\documentclass[useAMS,usenatbib]{mn2e}
\usepackage{graphicx}                                            
\usepackage{times}
\usepackage{float}
\usepackage{rotating}
\usepackage{epstopdf}
\usepackage{multirow}
\usepackage{hyperref}
\usepackage{color}
\def\lesssim{\mathrel{\hbox{\rlap{\hbox{\lower4pt\hbox{$\sim$}}}\hbox{$<$}}}}
\let\la=\lesssim
\def\gtrsim{\mathrel{\hbox{\rlap{\hbox{\lower4pt\hbox{$\sim$}}}\hbox{$>$}}}}
\let\ga=\gtrsim

\def\lx{$L_\mathrm{X}$}

\def\ergs{erg s$^{-1}$}
\def\ledd{$L_\mathrm{Edd}$}

\title[Black hole-like hysteresis in neutron star LMXBs]{Black hole-like hysteresis and accretion states in neutron star\\ low mass X-ray binaries}
\author[Mu\~noz-Darias et al.]{T.~Mu\~noz-Darias$^{1,2}$, R.~P.~Fender$^{1,2}$, S.~E.~Motta$^{3}$, T.~M.~Belloni$^{4}$\\
$^{1}$University of Oxford, Department of Physics, Astrophysics, Keble Road, Oxford, OX1 3RH, United Kingdom\\
$^{2}$University of Southampton, School of Physics and Astronomy, Southampton, Hampshire, SO17 1BJ, United Kingdom \\
$^{3}$ European Space Astronomy Centre (ESAC)/ESA, PO Box 78, E-28691 Villanueva de la Ca\~nada, Madrid, Spain\\
$^{4}$ INAF-Osservatorio Astronomico di Brera, Via E. Bianchi 46, I-23807 Merate (LC), Italy\\
}

\begin{document}
\maketitle
\begin{abstract}
We have systematically studied a large sample of the neutron star low mass X-ray binaries (LMXBs) monitored by the \textit{Rossi X-ray Timing Explorer} (50 sources; 10000+ observations). We find that the hysteresis patterns between Compton dominated and thermal dominated states, typically observed in black hole LMXBs, are also common in neutron star systems. These patterns, which also sample intermediate states, are found when looking at the evolution of both X-ray colour and fast variability of ten systems accreting below $\sim 30\%$ of the Eddington Luminosity (\ledd). We show that hysteresis does not require large changes in luminosity and it is the natural form that state transitions take at these luminosities. At higher accretion rates neutron stars do not show hysteresis, and they remain in a thermal dominated, low variability state, characterized by flaring behaviour and fast colour changes. Only at luminosities close to \ledd, are high variability levels seen again, in correspondence to an increase in the fractional contribution of the Comptonization component. We compare this behaviour with that observed in LMXBs harbouring black holes, showing that the spectral, timing and multi-wavelength properties of a given source can be determined by its location in the fast variability-luminosity diagram, which, therefore, provides a common framework for neutron star and black hole accretion states. \\      
\end{abstract}
\begin{keywords}
accretion, accretion discs, black hole physics, stars: neutron, stars: black holes, X-rays: binaries
\end{keywords}
\section{Introduction}
\label{intro}
Low mass X-ray binaries (LMXBs) are accreting sources harbouring a Sun-like star that is transferring matter on to a low magnetic field neutron star (NS) or a black hole (BH). Accretion takes place via an accretion disc, where matter is heated up to temperatures high enough ($\sim 10^7$ K) to radiate in the X-ray regime \citep{Shakura1973}. LMXBs can be divided in two populations according to their long term behaviour. Persistent sources remain active all the time, with X-ray luminosities typically above \lx $\sim 10^{36}$ \ergs. Contrastingly, transient systems spend most part of their lives in a dim (\lx $\sim 10^{30}$--$10^{34}$ \ergs), quiescent state, displaying occasional outbursts with recurrence times of months to decades, where \lx~ increases by several orders of magnitude. Apart from a couple of exceptions, BH-LMXBs belong to the latter group. Their study in the recent years has clearly established that they recurrently follow a similar evolution when they go into outburst, displaying  various X-ray states  possibly connected to different accretion regimes onto the BH (e.g. \citealt{Miyamoto1992},~1993;~\citealt*{Belloni2011}).  A $hard$ state is observed during the initial rise, when the energy spectrum is dominated by Comptonized emission and high levels of fast variability are seen. 
It precedes a fast transition (days) towards a $soft$ state at roughly constant luminosity (typically tens of \% of the Eddington Luminosity, \ledd). There, a thermal accretion disc component becomes dominant and fast variability disappears. The luminosity slowly (weeks-to-months) decays  down to a few per cent of \ledd\ during the soft state, after which a reverse transition to the hard state, also at fairly constant flux, occurs. This behaviour, firstly described by \citet{Miyamoto1995}, is hysteresis, since systems initially evolve from the hard to the soft state to then return to the former following a different luminosity track. Intermediate states, with very distinctive timing properties can be also identified during the two major state transitions (see also \citealt{Belloni2010}). The physical mechanism behind this hysteresis is still not well understood, but explanations related to the accretion disc magnetic field (\citealt{Balbus2008}, \citealt{Petrucci2008}, \citealt{Begelman2014}) or the Lense–Thirring effect \citep{Nixon2014} have been proposed.

Hysteresis loops are seen in most of the black hole transients (BHT; e.g. \citealt{Dunn2010}) when looking at the tracks that they display in the hardness-intensity diagram (HID, \citealt{Homan2001}; \citealt{Belloni2006}). The HID traces the evolution of the X-ray luminosity and energy spectrum along an outburst. Loops can have different shapes depending on the system (\citealt{Munoz-Darias2013b}) but share common properties like being always run in anti-clockwise direction. Also, the different branches of the HID are tightly correlated with the presence and type of outflows, both in the form of jets (best observed in the radio band; \citealt*{Fender2004}) or winds (\citealt{Ponti2012}). Recently, \cite*{Munoz-Darias2011} proved that hysteresis tracks are also found when studying the evolution of the fast aperiodic variability through the root-mean-square (rms)-intensity diagram (RID), their different branches being similarly connected with  the canonical X-ray states (see also Mu\~noz-Darias et al. 2011b, \citealt{Heil2012}).\par

\begin{table*}
\centering
\caption{Transient (top) and persistent (bottom) sources showing full hysteresis loops.}
\begin{tabular}{l c c c c c }
\hline
NS-LMXB & $N_{\rm H}$ ($\times 10^{22}$ cm$^{-2}$)    & Behaviour$^{~*}$ & Loop duration (d) & Hard-to-Soft times (d)$^{~**}$ & Soft-to-Hard times (d)$^{~**}$\\
\hline

\hline
EXO 0748-676 & 0.06 (1) & LTT, ML & 60--90 & $<$50 & $<$2~/~$<$8\\
Aql X-1 & 0.34 (2) & T, ML & 30--60 & $<$2~/~1--4 & $<$1 / $<$4\\
EXO 1745-278 & 1.2 (3) & T, SL & 90 & $<3$ &\\
4U 1608-52 & 1.6 (4) & T, ML & 25--90 &$<3$~/~2-6 & 2-4~/~$<7$\\
IGR 17473-2721 & 4 (5) & T, SL & 110 & 2--6 & 3--5\\ 

\hline

\hline
4U 1820-303 & 0.28 (4) & P, SL & $>56$ & $<$16 &\\
4U 0614+09 & 0.3 (4) & P, SL & 90 & $<13$ & $<23$ \\
4U 1636-53 & 0.43 (4) & P, ML &35--50 & $<2 $~/~ 2--6 & $<4 $~/~ 6--12\\
4U 1705-44 & 1.8 (6) & P, ML & 50--300 & $<4 $~/~$ < 8$ &4--16\\
4U 1728-34 & 2.5 (7) & P, ML & 20--60 & $<4$~/~2--6 & $<2$~/~ 4--8 \\

\hline
\end{tabular}
\begin{flushleft} 
\textit{*}~T: transient; LTT: long-term transient; P: persistent; ML: multiple loops; SL: single loop\\ 
\textit{**}~Minimum / maximum transition time scales are quoted if more than one loop is seen.\\
REFERENCES:  
(1) \cite{DiazTrigo2011}; (2) \cite{Maccarone2003a}; (3) \cite{Heinke2003}; (4) \cite{Christian1997}; (5) \cite{Chenevez2011}; (6) \cite{diSalvo2009}; \cite{diSalvo2000}

\end{flushleft}
\label{log}
\end{table*}

NS systems account for the vast majority of the persistent LMXBs, although they are also commonly found as transient sources. The X-ray phenomenology displayed by NS-LMXBs seems to be somehow more complex than for BH systems, probably due to the presence of a hard surface and anchored magnetic field. However, they also display several X-ray states, some of them analogous to those in BHs (see \citealt{vanderklis2006} for a review). At high accretion rate (\lx $\gtrsim 0.5$ \ledd) NS display Z shaped tracks both in the HID and in the colour-colour diagram (CCD), hence the name \textit{Z-sources}. Lower accretion rate systems  (0.01 \ledd $\lesssim$ \lx $\lesssim 0.5$ \ledd) show a different phenomenology and are classified as \textit{atoll sources} also based on the tracks displayed in the CCD \citep{Hasinger1989}. They show three main X-ray states, which can be somehow identified with  the hard, intermediate and soft BH states (e.g. \citealt{vanderklis2006}). A remarkable proof of the above scheme came from the study of the NS transient XTE J1701-462, which showed both Z and atoll phases along its decay from a bright outburst [see \citealt*{Lin2009a} (LRH09); \citealt{Homan2010}]. Hysteresis has been seen in a few atoll sources, and especially for the case of the transient system Aql X-1 clear analogies to BHTs have been reported (\citealt{Maccarone2003a}). However, up to now this behaviour was thought to be associated with transient sources (but see \citealt{Belloni2007}, 2010), and particularly to BHTs with some notorious NS exceptions (e.g. \citealt{vanderklis2006}, \citealt*{Gladstone2007}).\\
Here, we use archival \textit{Rossi X-ray timing Explorer} (RXTE) observations of NS-LMXBs to systematically search for hysteresis loops in both the spectral and the variability behaviour of these sources. We apply a blind BH-like approach by computing HIDs and RIDs for every source and show that hysteresis, at least qualitatively similar to that observed in BHs, is also a common feature in the persistent and the transient NS-LMXB population when accreting at medium rates. In section \ref{zetas} we show that hysteresis is suppressed at high accretion rates, but other patterns can be identified. In section \ref{discussion} we discuss the fact that systems located in the same RID area share similar phenomenology, yielding a common framework for BH and NS accretion states.

\section{The sample and data analysis}

We have analysed 50 NS-LMXBs densely monitored ($\gtrsim 20$ times) by RXTE (see Table \ref{tab:full-list}) using 10816 observations. We have included all the systems where type I X-ray bursts were detected (\citealt{Galloway2008}) plus other sources found in the literature. The vast majority of the systems are either transient or persistent NS-LMXBs accreting at low/moderate rates (i.e. atolls) but the sample also includes the population of Z-sources (see section \ref{zetas}) with the exception of Sco X-1, that was excluded from this global study due to its extremely high count rates, which result in a variety of non-standard observing modes. Systems within regions dense in X-ray sources (e.g. the galactic centre) that could not properly be singled out by RXTE were also excluded. We computed the RID and the HID for the full sample using data from the \textit{Proportional Counter Array} (PCA). The variability study was performed using a combination of high time resolution modes. A power density spectra (PDS) per observation was computed following the procedure outlined in \cite{Belloni2006}, stretches 16 s long and STD2 channels 0--31 (2--15 keV). The RIDs were computed following \cite{Munoz-Darias2011} and using the 0.1--64 Hz frequency band.\\
The RXTE-PCA Standard 2 mode (STD2) was used for the production of the HIDs. It covers the 2--60 keV energy range with 129 channels. For each observation net count rate (i.e. background subtracted) corresponds to STD2 channels 0--31 (2--15 keV). \textit{Hardness} was defined as the ratio of counts between the channels 20--33 (10--16 keV) and 11--19 (6--10 keV), respectively. This choice of bands excluding the softest channels is convenient for practical reasons since it minimizes the effect of interstellar absorption and avoids the more complex soft spectral region where the power-law, NS surface and thermal disc components coexist. For observations taken before 13 May 2005 the channel ranges were slightly shifted according to the corresponding RXTE gain epoch\footnote{\url{http://heasarc.gsfc.nasa.gov/docs/xte/e-c_table.html}}.  For each source, we analysed all the available data with count rates above 10 counts (per PCU). In a few cases we do not show all of them in order to e.g. skip some periods/gain epochs with low monitoring cadence that do not add any relevant information. For 4U~1636-536, we only present for clarity the high-cadence data already shown in \cite{Belloni2007} since they are fully representative of the behaviour of the source. 

\section{Results}
We have computed the HID and the RID for the 50 sources included in our sample (see Table \ref{tab:full-list}), finding a very distinctive behaviour between low (atolls) and high (Z) accretion rate systems. In particular we observe hysteresis in a significant fraction of the former, both persistent and transient objects (see Table \ref{log}), whilst is not observed at high luminosities.

\subsection{Low accretion rate sources}
Low accretion rate sources (0.01 -- 0.5 of \ledd) are known to display several X-ray states (see Section \ref{intro}) depending (but not exclusively) on their luminosity.  In our analysis we find a known population of relatively faint sources displaying exclusively high variability levels and hard colours typical of hard states (e.g. SAX~J1808.4-3658; see Tab. \ref{tab:full-list}), whereas the brightest atolls are always softer and much less variable (see section \ref{ba}). However, a significant fraction of the sample (21 out of 50) show both states, and whenever the RXTE cadence is good enough to follow their evolution we observe that the transitions between these states always happen following an hysteresis pattern. We observe full hysteresis cycles in ten systems (see Figs. \ref{h_tran} and \ref{h_per}), including four transients, one long-term transient (outburst lasts several years) and five persistent sources. In this paper we study the main properties of these hysteresis patterns and compare them with those found in BHTs. 

\subsubsection{Transient systems}
\begin{figure*}
\centering
\includegraphics[keepaspectratio]{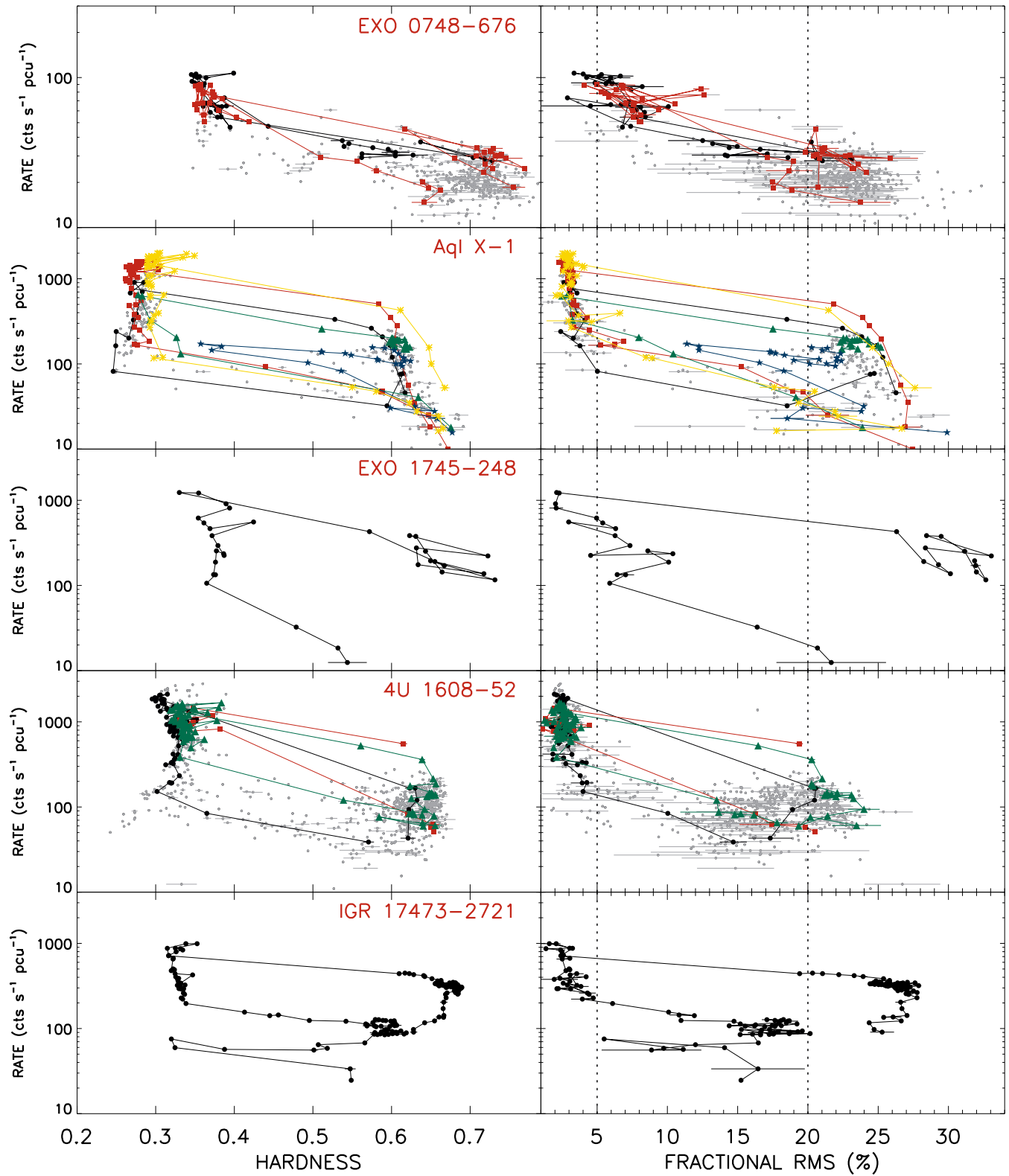}

\caption{Hysteresis in transient neutron star systems. Hardness intensity diagrams (left panels) and rms intensity diagrams (right panels) for transient sources showing hysteresis. Sources are sorted by increasing $N_{\rm H}$ from top to bottom. Solid lines join observations corresponding to full hysteresis cycles, which are highlighted with different symbols and colours. For the case of Aql X-1 only 5 representative loops are indicated. Error bars are shown every five points.}
\label{h_tran}
\end{figure*}

We observe hysteresis loops in the HIDs and RIDs of five transient NSs and they are shown in Fig. \ref{h_tran}. Systems are sorted by increasing neutral hydrogen column density along the line of sight ($N_{\rm H}$) as it is expected to affect the hardness (see section \ref{facts}). For all the systems but EXO~0748-676 each loop corresponds to one outburst. Points connected by a solid line correspond to the same hysteresis loop, whereas non-connected points correspond to observations which could not be included within a particular cycle. In most of the cases the latter is due to poor RXTE sampling or observations starting late in the outburst. The two loops found in EXO 0748-676 are observed within the same outburst that lasted more than 23 years and during which the source behaved as a persistent source. Indeed, these loops, associated with flares in the light curves with typical time scales of months ($\sim 60$ days in this case), are similar to those in persistent sources (see below). Hysteresis-like behaviour is seen in (at least) other three transients (see Tab. \ref{tab:full-list} ) but the RXTE monitoring was not intense enough to follow one full loop.\par

\subsubsection{Persistent systems}
\begin{figure*}
\centering
\includegraphics[keepaspectratio]{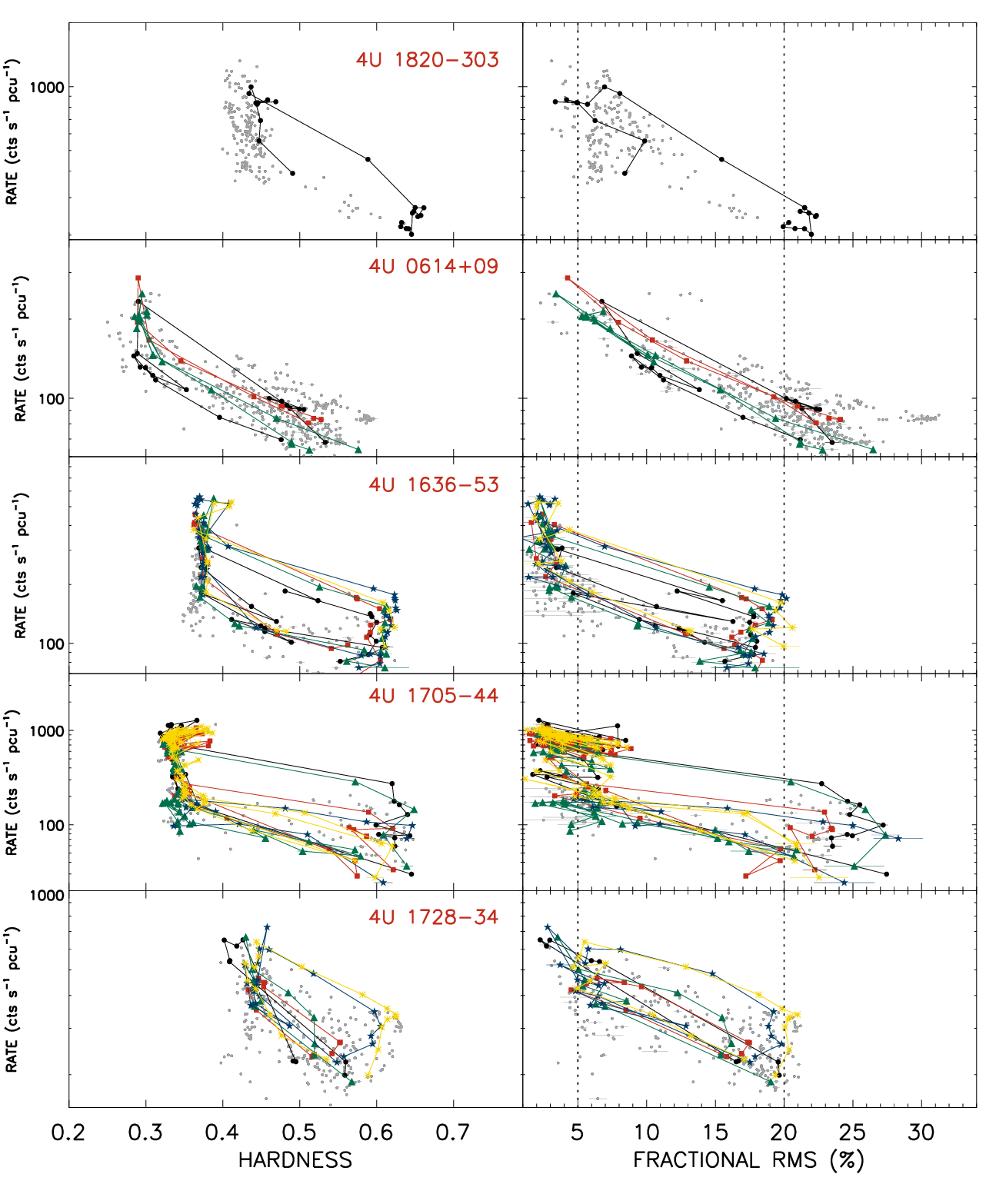}

\caption{Hysteresis in persistent neutron star systems. Same as Fig. \ref{h_tran} but for persistent NS-LMXBs. For the cases of 4U 1636-53, 4U 1705-44 and 4U 1728-34 only 5 representative loops are highlighted for clarity.}
\label{h_per}
\end{figure*}

Hysteresis loops are also found in five persistent systems (Fig. \ref{h_per}; also sorted by increasing $N_{\rm H}$) and at least other four seem to display similar behaviour (not shown; see Table \ref{tab:full-list}). As for the transients, each hysteresis episode is related to a flux increase (see Fig. \ref{persistent_lc}), although persistent sources do not decay towards quiescence once they are back to the hard state. The loop reported for 4U 1820-303 is not fully complete since observations stopped before the system reached again the hard state. However, the final hardening and rms increase together with the distribution of the remaining observations (grey points) strongly suggest that the loop was indeed completed. 

\subsubsection{Hysteresis loops: observational facts}
\label{facts}
\begin{figure}
\centering
\includegraphics[keepaspectratio]{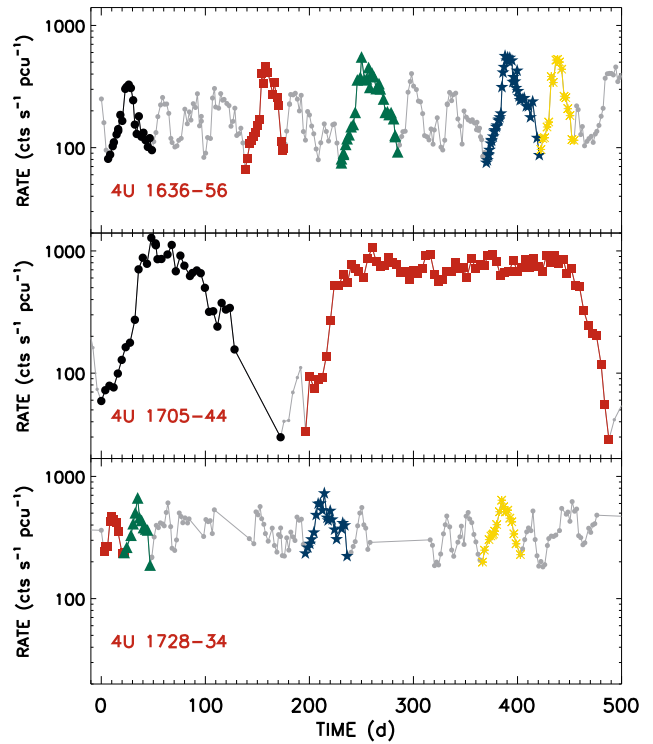}
\caption{Examples of light curves for 4U 1636-536, 4U 1705-44 and 4U 1728-34. We show 500 days intervals starting on MJD 53436, MJD 54296 and MJD 53798, respectively. Corresponding hysteresis loops are plotted using the same symbols and colours than in Fig. \ref{h_per}.}
\label{persistent_lc}
\end{figure}

The hysteresis loops observed in both transient and persistent sources share many properties. Sources are found to be hard and variable (fractional rms in the range 20--30 $\%$) during the rise (hard state) but, at some point, a transition to a softer and less variable state ($\leq 5 \%$) occurs. Indeed, the presence of a soft branch with 2-3\% rms is obvious in at least 6 of the sources. The luminosity eventually decreases during this soft state, which precedes a reverse (i.e. soft-to-hard) transition after which similar colours and variability levels than during the initial rise are newly displayed. Therefore, as in BHTs, these hysteresis loops are always run anti-clockwise. The range of hardness sampled (left panels in Figs. \ref{h_tran} and \ref{h_per}) is 0.25--0.75. The differences in $N_{\rm H}$ (see Tab. \ref{log}) do not seem able to explain the small differences in colours, even though $N_{\rm H}$ varies by an order of magnitude (excluding EXO 0748-676; see below). We note, however, that a clear trend with $N_{\rm H}$ is obtained (not shown) when using softer bands than those selected for the hardness. Variability patterns (i.e. RIDs; right panels in Fig. \ref{h_tran}) are very similar between each other, and also resemble those seen in BHT (Mu\~noz-Darias et al. 2011, 2011b, \citealt{Stiele2012}, \citealt{Motta2012}). 
Besides these general characteristics, we should also highlight the following observational facts:
\begin{itemize}
\item \textit{Diagonal state transitions:} the count-rate increases during the hard-to-soft transition and decreases during the reverse transition. In other words, the transitions are always diagonal and never parallel as it is the case for low inclination black holes using the same energy bands. Triangular shapes similar to those seen for high inclination black holes (\citealt{Munoz-Darias2013b}) do not seem to be present  in the eclipsing source EXO 0748-676. Therefore, unlike in the case of BHTs, sources are systematically brighter (sometimes by one order of magnitude) during the soft state than during the hard or the intermediate states.
\item \textit{Transition luminosities:} persistent systems show very similar soft-to-hard (i.e. lower branch) transition luminosities for a given source, whilst larger scatter is observed for the upper branch. The case of 4U 1705-44 is particularly interesting  since the count rate of the hard-to-soft transition varies from levels a factor of $\sim 2$ to one order of magnitude brighter than the lower branches. In particular, one of its loops (yellow asterisks) seems particularly complex. After performing several faint loops ($\lesssim 200$ cts s$^{-1}$; e.g. blue stars) the system displayed a new faint hard-to-soft transition and almost completed the loop with a \textit{hard excursion}. However, instead of continuing to harden, it apparently (with gap between observations of just 4 days) returned to the soft state to then became one order of magnitude brighter. Subsequently, a decay and a final soft-to-hard transition at a count-rate higher than the previous hard excursion, but still slightly fainter than the initial upper branch, was observed. An alternative explanation is that we were looking at two different loops, the initial part of the second being faster than 4 days.\\
Aql X-1 is the only transient with several, well monitored outbursts, and lower branch count-rates cluster within a factor of $\sim 2$ [$\sim 3$ if we consider the failed outburst (see below) shown as blue filled stars in Fig. \ref{h_tran}]. On the other hand, the two best sampled loops of 4U 1608-52 (dots and triangles) show lower branch count-rates within a factor of $\sim 3$ (the outburst represented by squares has a one week gap during the soft-to-hard transition).
\item \textit{Time scales:} estimations of the loop time scales are shown in table \ref{log}. Loop durations are measured from the time the system starts to continuously rise along the hard state until it gets back to it after a hysteresis cycle. These times should be taken as approximations with typical errors of  $\sim$ 5 days. Loops last $\sim$ 1 to 4 months, a particular source being able to show different (factor of 2) cycle durations. Loop times for persistent sources are consistent with those for the transients, with the exception of 4U 1705-44 with three cycles lasting $\sim$ 170, 290 and 300 days (see Fig \ref{persistent_lc}). These atypically long loops are due to long periods with the system in a bright soft state; however, transition times are consistent with those of the other NS-LMXBs. 

Transition time scales are more difficult to estimate since they are too short to be properly sampled given the $\sim$ 1--2 days cadence (at best) of the RXTE monitoring. Hard-to-soft and soft-to-hard transition times seem consistent between each other and vary from $\lesssim 1$ to $\sim 6$ days. The fact that fewer observations seem to happen during the former than during the latter transition is, for the transients, mainly a consequence of observations starting late in the outburst (soft state) as already pointed out by \cite{Tudose2009} for the case of Aql X-1. Moreover, transition luminosities (see above) are more similar during the lower than during the upper branch, contributing to cluster points around count-rate levels corresponding to the former.
\end{itemize}

Finally, we note that four sources (4U 1820-303, 4U 1728-34 and 4U 0614+09 and EXO 0748-676) show variability levels rarely below 5 \%, the first two also showing harder soft state colours than the others.  4U 0614+09 manifests a peculiar behaviour, with only one of the three loops showing clear hysteresis (filled triangles). In the other two tracks showed, the system describes a single path, with transitions occurring at similar flux levels (although the hard-to-soft branches are still slightly brighter). The case of EXO 0748-676 is also peculiar. It is seen through very low interstellar extinction but observations suffer from the intrinsic eclipsing and dipping behaviour of the source due to its high inclination, which are responsible for the large scatter present in the data. However, we note that several soft observations do not show either dips or eclipses and still look harder than for the other sources. Another peculiarity is the \textit{failed transition} that seems to be also present in at least one of the outburst of Aql X-1 (green stars; Fig. \ref{h_tran} second panel). Outbursts with failed transitions (see \citealt{Capitanio2009}, \citealt*{Motta2010}, \citealt{Soleri2013} for examples in BHT), are essentially only-hard state outbursts with one or multiple softer intervals (see also \citealt{shaw2013}) where the accretion disc component never becomes dominant, and the timing and spectral properties never go beyond those corresponding to the intermediate states.

\subsubsection{The hardness-rms correlation}
\label{hrd}
The hardness-rms diagram (HRD) is often used to study the outburst evolution of transient LMXBs, especially those with BH accretors (see e.g. \citealt{Belloni2005}, \citealt*{Fender2009}). Besides some features regarding detailed state classification, the most basic property that becomes apparent in the HRD is a positive correlation between hardness and rms observed across the whole outburst. Sources displaying hysteresis in the HID and the RID (i.e. those presented in Figs. \ref{h_tran} and \ref{h_per}) do not show it in the HRD, where the correlation is the same for both the upper and lower branches. In Fig. \ref{hrd_plot} we show the HRD for two representative sources from our sample, namely, Aql X-1 (transient) and 4U 1705-44 (persistent), where the aforementioned correlation is evident. In addition, we also observe that when systems reach variability levels $\lesssim 10$\%, the hardness remains constant and variability drops, and then the latter stays the same ($\sim$ 2-3 \%) as hardness increases, yielding a hook shape. It is formed by the highest count-rate observations, and for the case of 4U 1705-44, can last several months (see also Fig. \ref{persistent_lc}). These hook-like tracks are not observed in BH systems and seem to be a distinctive feature of NS accretors.

\subsubsection{Bright Atolls}
\label{ba}
We have shown that hysteresis is present in some persistent, low accretion rate sources. However, there is also a population of (persistent) sources that remain almost permanently in the bright soft states; they are known as \textit{bright atolls} (BA). In Fig. \ref{hrd_plot} we plot the HRD for two of these systems,  4U 1735-444 and GX~9+9. The former only displays the hook-like track mentioned above and characteristic of the bright soft states, whereas the latter, one of the most luminous atolls, was observed 70 times from 1996 to 2010 and all the observations show constant 2--3\% rms as hardness moves in the ranges 0.3--0.42 and count rate changes by a factor of $\lesssim 2$ (see Fig. \ref{hrd_plot}). Other examples of bright atoll sources are GX 3+1, GX 13+1, GX 9+1 or GX 9+9 (see Tab. \ref{tab:full-list}) and they never show variability levels above 10\%.

\subsection{High accretion rate sources}
\label{zetas}
At even higher accretion rates than those of the bright atolls, NS-LMXBs are known to display a particular short-term behaviour (hours-to-days) in their CCDs and HIDs. They are so-called Z sources since they show three connected and distinctive branches in the CCD, known as the flaring branch (FB), the normal branch (NB) and the horizontal branch (HB), all together forming a Z shaped pattern (see \citealt{vanderklis2006}). They can be divided in two sub-groups, Cyg-like and Sco-like sources, the latter not showing a proper HB \citep{Kuulkers1994} and thought to be intrinsically fainter \citep{Homan2007}. Their long term behaviour is however more erratic and complex than the low accretion rate sources presented above. We have analysed the RXTE observations corresponding to eight sources that have shown Z-like phenomenology, namely Cyg X-2, GX 5-1, GX 17+2, GX 340+00, GX 349+2, XTE J1701-462 and LMC X-2 (see Tab. \ref{tab:full-list}). Similarly to bright atolls, hysteresis cycles are not seen in these systems. In the upper panels of Fig. \ref{fig:zetas} we show the HIDs and RIDs of two representative objects: Cyg X-2 (i.e. Cyg-like) and GX 17+2 (Sco-like). 

\subsubsection{The case of XTE~J1701-462}
\begin{figure*}
\centering
\includegraphics[keepaspectratio]{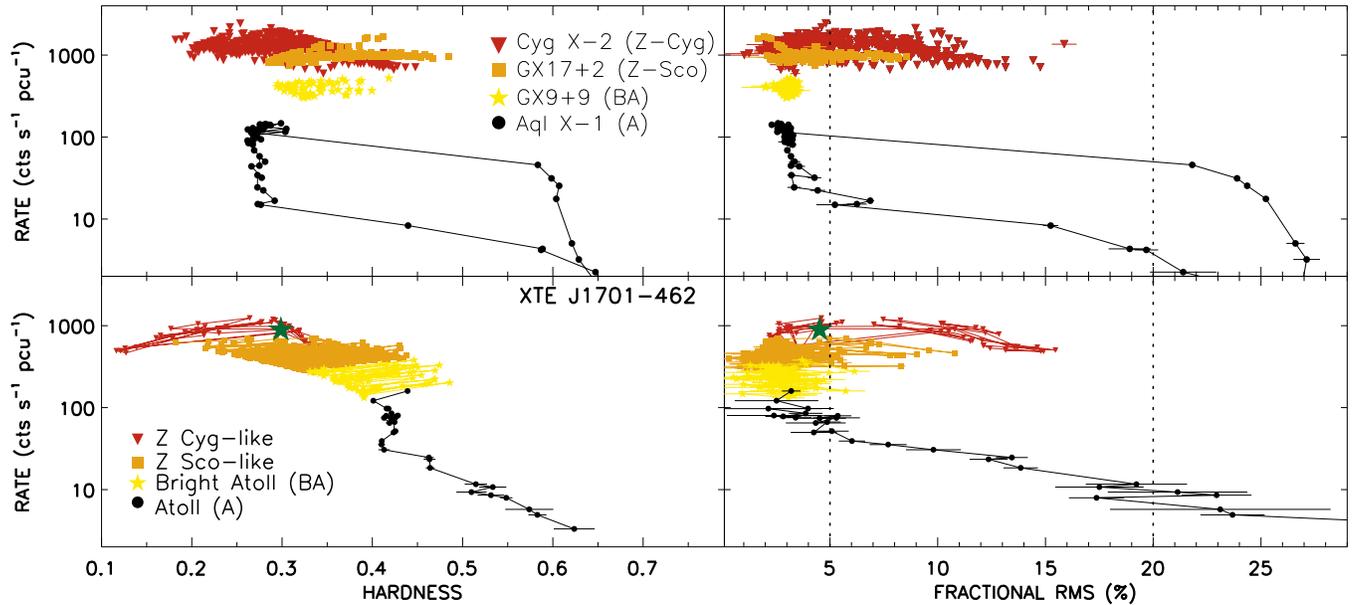}

\caption{Hardness intensity diagrams (left panels) and rms intensity diagrams (right panels) for the Z sources  Cyg X-2 and GX 17+2, the bright atoll source GX 9+9 and the transient Aql X-1 (upper panels). In the bottom panels the HID and RID corresponding to XTE~J1701-462 are shown. For the latter, symbols and colours correspond to the outburst stages defined in \citet{Lin2009a} and the (green) star indicates the first observation. For a more comprehensive picture, all count rates were re-scaled to that of Cyg X-2. We use 13 kpc for Cyg X-2 (the reported value is 11$\pm$2 kpc. 13 kpc is chosen since it qualitatively produces a better match between the two Z sources in the sense that their combined HID behaviour better mimic that seen for XTE J1701-462), 9.8 kpc for GX 17+2, 3.9 kpc for Aql X-1 (\citealt{Galloway2008}), 10 kpc for GX 9+9 \citep{Savolainen2009} and 8.8 kpc for  XTE~J1701-462 \citep{Lin2009}.}
\label{fig:zetas}
\end{figure*}

\begin{figure}
\centering
\includegraphics[keepaspectratio]{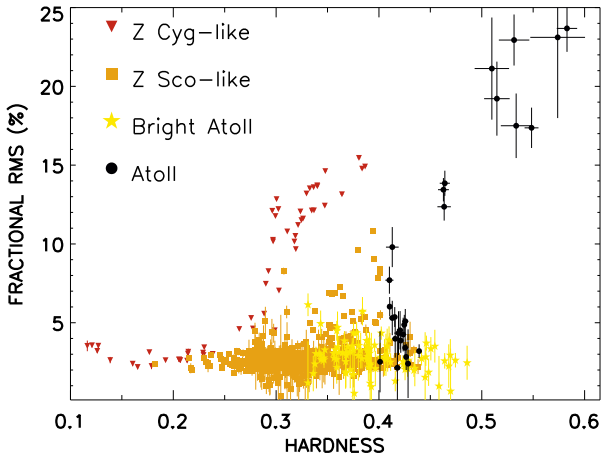}

\caption{Hardness rms diagram for XTE J1701-462. Colours and symbols represent the same as in Fig. \ref{fig:zetas}.}
\label{fig:hrd_1701}
\end{figure}

\begin{figure}
\centering
\includegraphics[width=9cm, height=6cm]{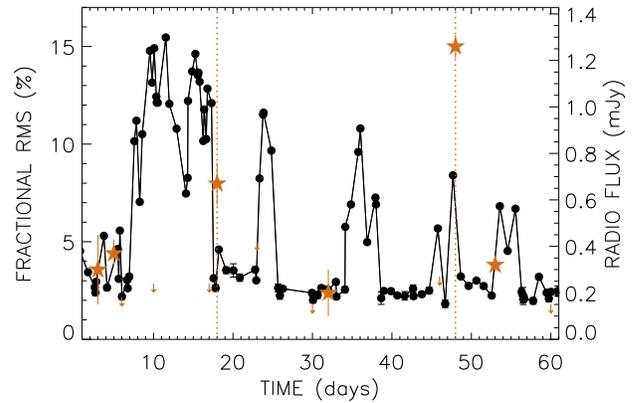}
\caption{Fractional rms (black dots) and radio data (orange stars; from \citealt{Fender2007}) as a function of the time corresponding to the 2006 outburst of XTE J1701-462. The arrows indicate $3\sigma$ limits to the radio flux. The zero time corresponds to MJD 53754.}
\label{fig:radio}
\end{figure}

A unique system to study the behaviour of NSs at different accretion rates is the transient XTE~J1701-462. This source displayed both Z and atoll phenomenology along its decay from a bright outburst (see \citealt{Homan2007}, LRH09, Lin et al. 2009b, \citealt{Sanna2010}, \citealt{Homan2010} for details). Its corresponding HID and RID are shown in the lower panels of Fig. \ref{fig:zetas}. Following LRH09, we separate (using different symbols and colours) the Z (Cyg-like and Sco-like), bright atoll and normal atoll phases. In Fig. \ref{fig:hrd_1701} we do the same for the HRD. The initial rise of XTE J1701-462 was missed, as it was already very bright during the first RXTE observation (big star in Fig. \ref{fig:zetas}), moving through the Cyg-like Z regime. During the first week, it displayed low variability levels ($\sim3\%$), similar to the soft states of atoll sources, but with softer colours typical of Z-sources. Then, the system moved to a more variable (7--15\%; see Fig. \ref{fig:radio}) and harder state (i.e. hardness and rms are correlated; see red triangles in Fig. \ref{fig:hrd_1701}), where it stayed for $\sim$ 10 days. LRH09 showed that the Z phase observations with variability levels above $\sim 5\%$ are associated with the HB. This increase in the aperiodic variability when the system enters the HB have been also observed in other Z sources (e.g. see \citealt{Wijnands1997} and \citealt{Li2012} for Cyg X-2). During this phase,  XTE J1701-462 was intensively monitored, and the PDS of the corresponding observations are qualitatively similar to those of BHTs during the hard-to-soft transitions (see Section \ref{qpos}), for which the rms values ( $\gtrsim 5\%$) are also typical (\citealt{Munoz-Darias2011}). Similarly to what is observed in BHTs, the characteristic PDS frequencies increase as the system gets softer (\citealt{Homan2007}) and less variable. On Jan 5, 2006 (day 17 in Fig. \ref{fig:radio}), variability dropped from 12\% to 3\% within the $\sim 4$ hours that separate two consecutive RXTE observations (91106-02-03-12 and 91106-02-03-15 ), returning to a softer, low variability state. Interestingly, \citet{Fender2007} reported that radio emission (4.8 GHz) increased from  $< 0.24$ mJy\footnote{The limits to the radio detections correspond to 3 times the rms of the noise.} two hours before the transition, to $0.67\pm0.09$ mJy 40 hours later (see Fig. \ref{fig:radio}). After this, the system remained most of the time in the low variability state, with a handful of shorter extra excursions like the previous one. On Mar 8, the brightest radio flare observed from this source occurs ($1.26 \pm 0.5$ mJy; $< 0.27$ mJy two days before; Fig. \ref{fig:radio}) and rms dropped from  8\% on Mar 7 to 3\% one day later. We note that similar radio flares are also detected in BHT when moving between similar variability states during the hard-to-soft transition (see section \ref{radio}). 

\begin{figure}
\centering
\includegraphics[keepaspectratio]{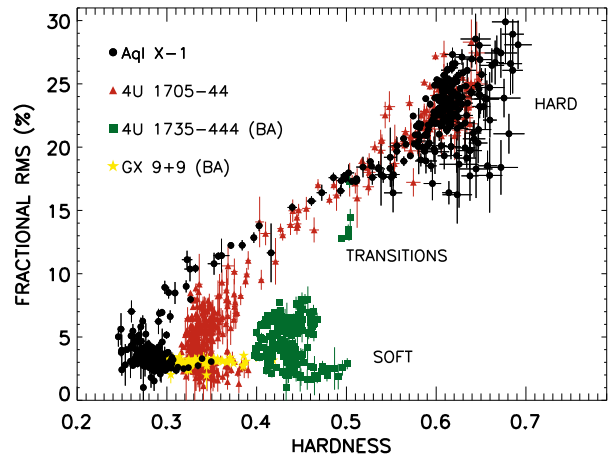}
\caption{Hardness-rms diagrams for Aql X-1 (transient) and 4U 1705-44 (persistent), where the rms-hardness correlation become evident. For comparison we also show those corresponding to the bright atolls (see section \ref{ba}) 4U 1735-444 and GX 9+9.}
\label{hrd_plot}
\end{figure}

As flux decays, these higher rms excursions are less frequent and the system remained in the low variability state all the time (rms $<5$\%). However, frequent changes in hardness are seen, the X-ray colour reaching their hardest values apart from the final hardening during the normal atoll phase. This period correspond to the faintest part of the Z Sco-like stage and the whole bright atoll phase, where fast variability and colour are not correlated and the horizontal track is formed in the HRD (Fig. \ref{fig:hrd_1701} and Section \ref{hrd}). This behaviour is suppressed once the systems enters into the regular atoll phase (filled dots in Fig. \ref{fig:zetas}). There, the system keeps decaying at a roughly constant colour and low variability levels along the atoll soft state, until a transition is observed, and the system hardens and rms increases above 20\%, typical of a standard hard state (rms and hardness correlated; Fig. \ref{fig:hrd_1701}). Hence, the RID pattern displayed by XTE J1701-462 is remarkably similar to those seen in the transient sources (Fig. \ref{h_tran}) once they reach a bright soft state. Based on the behaviour seen in other sources, it seems reasonable to speculate that during the initial rise / atoll phase (not observed) the system transited along the hard state and made the hard-to-soft transition, probably still during the atoll phase. After that, it became much brighter as seen in e.g. 4U~1705-44, but reaching the Z regime in this case. We note that a regular atoll hard state and subsequent hard-to-soft transition had been difficult to detect by the all-sky monitor on-board RXTE, given the soft pass band of the instrument and the relatively large distance to the source (8.8 kpc; \citealt{Lin2009}).

Finally, we have added the bright atoll source GX 9+9 and the transient Aql X-1 to the Z sources presented in the uppers panels of the Fig.\ref{fig:zetas}. It becomes evident that the behaviour of XTE J1701-462 in the RID and HID it is very similar to the combination of atolls, bright atolls and Z sources. \\

\section{Discussion}
\label{discussion}

We have analysed a large data base of RXTE observations ($> 10000$) corresponding to 50 NS-LMXBs. For each source, we have studied the behaviour of the spectrum and the fast aperiodic variability through the HID and the RID. Hysteresis loops have been observed, sometimes recurrently, in ten sources. Every source included in the sample has been monitored at least 20 times. However, these pointings are sometimes spread over several months/years, preventing a follow-up along a full hysteresis loop. We note that the behaviour of at least seven other sources suggest hysteresis (see Tab. \ref{tab:full-list}) and some transients, discovered being (already) soft, displayed the soft-to-hard transition when decaying towards quiescence (e.g. XTE J1701-462). In general, hysteresis occurs at low accretion rates whenever sources display full atoll-like behaviour, regardless if they are transient or persistent sources. Classical transients, like Aql-X-1, show the classical hysteresis previously seen in BHT, where one loop per outburst is observed. This is not the case for the long-term transients, that instead of showing a single hysteresis loop lasting several years (e.g. 23 yr for EXO 0748-676)  show also loops lasting a few months. They are associated with flux increases where the soft state is reached and after which the source returns to the initial hard state but does not decay towards the quiescence. This behaviour is similar to that seen in persistent sources. The above suggest that hysteresis loops are short by nature, as we do not see any loop lasting more than $\sim$ 1 year, consistent with what it is observed in BHTs.
Overall, we find a consistent behaviour between the loops properties, including the time scales, for both the transient and the persistent populations. Loop amplitudes (count-rate difference between upper and lower branches) are generally larger in the transients, but consistent with e.g. some of those seen in the persistent source 4U 1705-44. The range of loop amplitudes present in persistent systems goes from less than a factor of 2 to one order of magnitude. In this regard, our study (Figs. \ref{h_per} and \ref{persistent_lc}) proves that big changes in luminosity are not required to observe hysteresis. A flux increase of a factor of $\leq 3$ in a persistent source (e.g. 4U 1728-34) yield hysteresis loops covering a range of colours and variability levels comparable to those of the transients, where major changes in luminosity occur. For the case of 4U 0614+09 the two branches almost overlap in luminosity. We note, however, that the two major state transitions seem to occur at more similar luminosities during failed outburst (see section \ref{facts}) than during normal ones in BHTs and in Aql X-1 (this work). For 4U 0614+09 variability rarely go below $5\%$ and an obvious soft state branch is not present. This behaviour also resembles that of the persistent black hole Cyg X-1, where transitions from and to intermediate (i.e. softer but never reaching the soft state as defined for BHT) states are seen at roughly constant luminosity (e.g. \citealt{Wilms2006}).

\subsection{Hysteresis: black holes versus neutron stars}
\label{bhvsns}
\begin{figure}
\centering
\includegraphics[keepaspectratio]{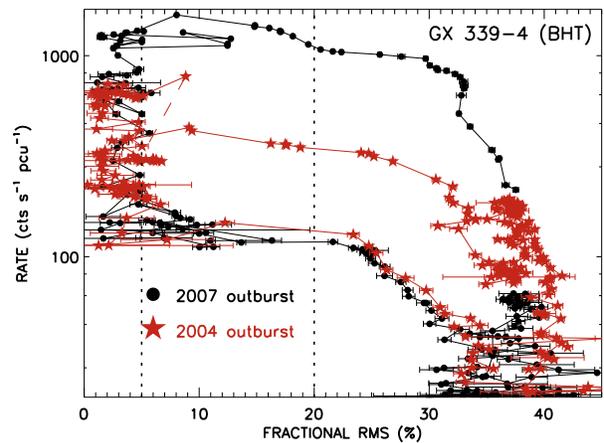}
\caption{RID for two outburst of the black hole transient GX 339-4 (adapted from Fig. 2 in \citealt{Munoz-Darias2011})}
\label{fig:gx339}
\end{figure}

Given that hysteresis has been traditionally associated with black hole transients, it seems natural to compare our results. A few conclusions can be directly extracted from our work:
\begin{itemize}
\item we have shown that the hysteresis in NSs follows the same pattern as in BHTs. The loops are always run anticlockwise, the hard-to-soft transition being the brightest. 
\item The luminosity of the upper branch changes along several loops from the same source (see Figs. \ref{h_tran} and \ref{h_per}; Fig. \ref{fig:gx339} for the BHT GX 339-4), whereas the soft-to-hard transitions, always fainter, seen to have more similar luminosities  (especially for the persistent systems), but still might vary within a factor of 2-3. The luminosity of the soft-to-hard transition for both BHs and NSs LMXBs have been proposed, using some of the sources presented here, to be always within 1-4\% of \ledd (\citealt{Maccarone2003}). Our study lacks spectral modelling and therefore is not able to test this claim, but our results seem consistent with it.
\item Hysteresis cycles duration times are consistent between BHs and NSs, roughly in the range  one month -- one year.
\item Transition times are poorly constrained by our study since they are typically of the order of our monitoring cadence. However, it is worth noticing that all the systems where we have decent constraints seem to point towards values shorter than a week. In BHTs transitions times longer than 10 days are quite common. 
Possible explanation for this might include different distribution of binary parameters (e.g. disc sizes: different mass ratios and orbital periods) or more complex relations between transition durations and e.g. compact object mass or luminosity that we are not able to test in this work. \par
\end{itemize}

We note that the several models have been proposed to explain the hysteresis cycles seen in BHTs (e.g., \citealt{Balbus2008}, \citealt{Petrucci2008}, \citealt{Nixon2014}, \citealt{Begelman2014}). Since the behaviour displayed by NS-LMXBs is very similar, it is therefore expected that hysteresis is triggered by the same physical mechanism in both cases. Hence, any model aimed at explaining this process should be also able to reproduce the observables reported in this work.   

\subsubsection{Three spectral components model for NS-LMXBs}
Further analysis of the results presented here requires consideration of the different spectral properties observed in BHs and NSs. The former have two distinctive spectral components: (i) a hard energy tail, sometimes extending beyond $\sim 100$ keV (e.g., \citealt{Motta2009}; \citealt{Gilfanov2010} for a review) that can be modelled (at low energies at least) by a power-law with a cutoff at energies higher than those considered here and is usually explained through Comptonization processes, and (ii) a soft thermal component (with temperature peaking at $\lesssim$ 2 keV) that is well described by a multi-colour disc spectral model. Both components are also present in NS-LMXBs. In addition, emission from the NS surface (i.e. absent in BHs) is also expected to contribute, and for the particular case of the inner disc reaching the innermost stable circular orbit, the radiation emitted from the boundary layer could be at least comparable with that of the latter (\citealt{Syunyaev1986}). This component typically has a higher characteristic temperature than the disc (see below), which can cause degeneracies in the spectral fitting (e.g. \citealt{Barret2001}). As a result of this, a variety of spectral models have been proposed (see \citealt{White1988} and \citealt{Mitsuda1989} for alternative scenarios). They essentially differ in the treatment of the boundary layer, which is modelled with either a single black-body or a Comptonized black-body, the latter accounting for part or the whole of the high energy emission. In general, a two spectral component approach has been utilized to model the spectrum of NS-LMXBs, however the particular components used change depending on accretion state and luminosity.\par  
Lin, Remillard \& Homan (2007; hereafter LRH07) have proposed a three component approach, which is able to describe the spectrum of NS-LMXBs from the atoll sources states (Aql X-1 and 4U 1608-52) to the brightest Z phases of  XTE J1701-462 (LRH09). They consider two thermal components [multi-colour accretion disc and black body emission form the neutron star (BB)] plus Comptonized emission modelled by a broken power-law component. Likewise, \cite{Done2003} found that the energy spectra of NS-LMXBs are similar to those of the BH systems if one considers the additional emission from the boundary layer together with the classical thermal disc and Comptonization components. Some of the results presented in this paper can be interpreted according to this scenario: 

\begin{itemize}
\item although a detailed study would require spectral fitting to obtain and compare unabsorbed fluxes, none of our atoll sources reaches the low hardness values observed with RXTE/PCA in the same bands for several BHTs (e.g. \citealt{Casella2004}; \citealt{Stiele2011}, \citealt{Motta2012}). These harder soft state colours can be explained by the contribution of the boundary layer component, which systematically peaks at harder energies than the disc component \footnote{$kT_{Disc}<2.0$ keV whereas $kT_{BB}$ can reach $\sim 3.0$ keV; $k$ is the Boltzmann constant; $T_{Disc}$ and $T_{BB}$ correspond to the temperature at the inner disc radius and the colour temperature, respectively (LRH09).}. 

\begin{figure*}
\centering
\includegraphics[width=16cm,height=10cm]{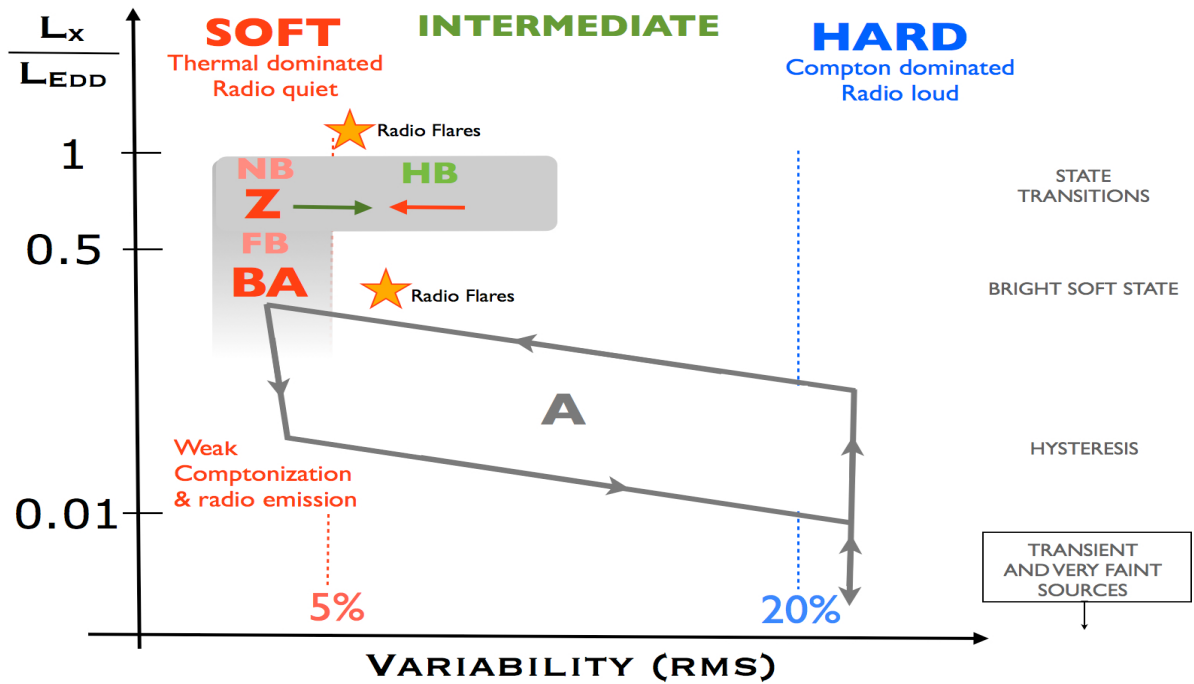}
\caption{Sketch describing the qualitative behaviour and location of atoll (A), bright atoll (BA) and zeta (Z) sources in the rms-intensity diagram.}
\label{fig:sketch}
\end{figure*}

\item LRH07 have shown that the three-component spectral model is able to reproduce for atoll systems (Aql X-1 and 4U 1608-52) the correlation between Comptonization fraction, and variability (rms) observed in BHTs (e.g. \citealt{Remillard2006}; \citealt{Belloni2011}). On the other hand, several works have proved in the past the remarkably similar timing properties of BH and NS systems, suggesting that the mechanism producing the variability are the same for both populations (e.g. \citealt{Wijnands1999a}; \citealt*{Belloni2002}). In the NS hard states variability reaches 20\% and in some cases 30\%, but is never seen to attain the $>40\%$ levels often seen in the hard states of some BHT (e.g. \citealt{Munoz-Darias2010}; Fig \ref{fig:gx339}). LRH07 showed that, even though the disc component is not required in atoll hard states, the BB component (i.e. also thermal) accounts for $\sim 10\%$ of the emission. Assuming that variability is only coming from the Comptonized component (the only component present in BHT hard states within the RXTE bandpass; e.g. \citealt{Dunn2010}), this additional thermal emission present in NS-LMXBs would dilute the variability by the same amount. Nevertheless, we note that a number of BHs have never showed rms above $\sim 30\%$ (e.g., 4U~1650-500, \citealt{Gierlinski1999}; H~1743-322, \citealt{Motta2010}). 

\item Soft state variability levels ($2-3$\%) are also similar to those found in BHTs, although levels as low as $<1\%$ do not seem to be present either. The fits presented in LHR07-09 suggest that Comptonization can still contribute to the emission during the soft states, especially at low luminosities.  We note that for a few systems (including those analysed by LRH07) soft state variability seems to increase as the systems get fainter. This weak Comptonization could also explain the radio properties of these systems during soft states (see section \ref{radio}).    
\end{itemize}

\subsection{The variability-luminosity diagram and the global accretion state picture}
In this section we discuss the evolution of NS-LMXBs (both atoll and Z sources) using the RID, and compare it with the spectral behaviour reported in this work (colours) and specially in the literature (e.g. LRH07, LRH09). The goal is to discuss the evolution of the aperiodic variability (rms) in NS systems, in light of the contribution of thermal (accretion disc + NS surface) and Comptonization processes (regardless of its origin) to their emission. Given the relation between variability and Comptonization discussed above, the RIDs presented in Fig. \ref{fig:zetas} for XTE J1701-462 and for a combination of sources displaying the different NS-LMXB X-ray states can be directly compared to those of BHTs.  

\begin{figure}
\centering
\includegraphics[width=10cm,keepaspectratio]{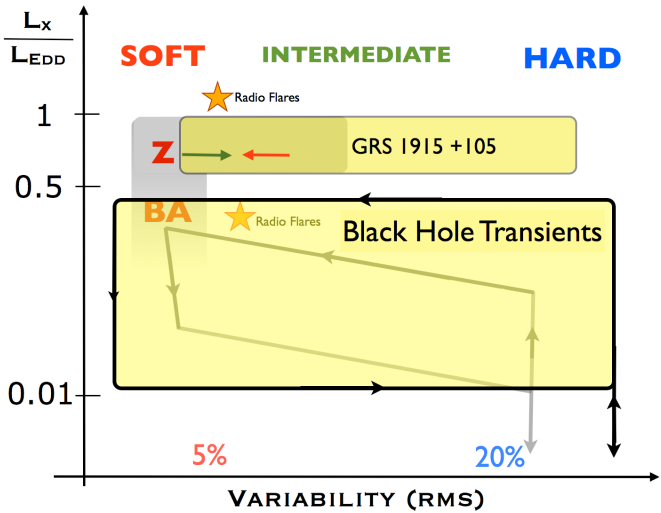}
\caption{Same as Fig. \ref{fig:sketch} but also adding the (qualitative) behaviour of black hole transients and GRS 1915+105. The latter is a persistent black hole that usually sits in the bright hard states. However, It occasionally samples softer regions, overlapping with those typical for the Z sources during their harder intervals (HB). Both state transitions yield strong jet emission.}
\label{fig:sketch_BH}
\end{figure}

In Fig \ref{fig:sketch} we sketch the behaviour of NS-LMXBs in the RID. The quoted luminosities have to be taken just as standard values typically found in the literature (e.g. \citealt{vanderklis2006}). In Fig. \ref{fig:sketch_BH} we do the same but we also over-plot the behaviour of BHTs (taking GX 339-4 as the standard source) for comparison. Looking at the variability axis, the diagram has two main regions, the hard state, where Comptonization dominates and strong aperiodic variability is present ($>20\%$), and the thermal dominated (accretion disc + NS surface) soft state with very little variability ($<5\%$). However, the way (frequency, duration, multi-wavelength properties) in which these regions are sampled by a given system seems to be strongly related (but not exclusively) to its luminosity (i.e. accretion rate). We distinguish three clear regimes\footnote{Note that the luminosity boundaries quoted are approximate and that transitions between different regimes can be smooth.} :
\begin{enumerate}

\item \textit{Hysteresis regime} (0.01 -- 0.3 \ledd)\\

At low luminosities sources display hysteresis loops very similar to those observed in BHTs at roughly the same brightness (e.g. \citealt{Dunn2010}; Fig. \ref{fig:sketch_BH}). This phase has already been extensively discussed in this paper (see sections \ref{facts} and \ref{bhvsns}) and it is also characterized by a clear correlation between fast variability (rms) and Comptonization fraction (hardness) as shown in Fig. \ref{hrd_plot}. Above this level, sources sit typically in the soft, low variability state, first as bright atoll sources (strictly always in the soft state), and then as Z sources.

We note that this regime (at least its lowest extreme) is also reached by systems (mostly transients) that do not go beyond the hard state, a behaviour also observed in BHTs. Also, and in addition to transients rising/decaying through the hard state, a new population of persistent NS-LMXBs intrinsically fainter than $\sim 0.01$ \ledd\ has recently been found (see \citealt*{ArmasPadilla2013}), even though not well understood yet.\\     

\item \textit{Bright soft state regime} (0.3 -- 0.6 \ledd)\\

This regime corresponds to the bright atoll and the faint Z phases, where the spectrum is always dominated by the thermal components, very little variability (2--3\%) is seen and flaring is present. As discussed above, during the regular atoll phase both variability and hardness are correlated, and both seem to be tracing the amount of comptonized emission. The correlation breaks during the bright atoll phases, when hardness varies but fast variability remains constantly low. This produces a clear signature in the HRDs, the normal atoll sources showing a hook shape, whereas bright atolls just show a flat rms-hardness relation (Fig. \ref{hrd_plot}). This pattern is unique to neutron star systems, and keeps being observed as luminosity increases up to the brightest Z phases (see below), where hardness and variability are newly correlated. Despite on the gradual softening as the system become more luminous (bottom-left panel in Fig. \ref{fig:zetas}) and the disc component become more dominant (LRH09), we do not see changes in HID, RID, and HRD during this stage. 

The above scenario is in agreement with LRH09, whose spectral fits clearly connect the bright atoll phases and the zeta FB phase of XTE J1701-462. They also show how, as luminosity increases, the systems starts to transverse the NB and even in the brightest part of the Sco-like phase (stage 2 in LRH09) it spent most of the time ($\sim 90\%$) between the FB and NB phases. LRH09 explain these two branches by (i) instabilities associated with the inner disc radius moving outward from the atoll soft state value and inward again (FB), and (ii) changes in the radius of the BB component (NB). Other authors (e.g. \citealt{Church2006}) suggest that the flaring is the result of unstable nuclear burning on the neutron star surface. 

We note that this phase, where hardness (energy spectrum) varies whereas the fast variability properties remain essentially the same, has not been seen in BHs, suggesting it has something to do with the presence of a hard surface. Nevertheless, it does not seem to be associated with Comptonization, explaining the lack of variability (always below $\sim 5$) during this phase, and therefore should be classified as a soft state phase.\\

\item \textit{Very high accretion rate regime: soft state and fast transitions} ($\gtrsim$0.6\ledd)\\

This regime corresponds to the brightest Z phases, where the spectrum is typically soft and the level of variability very low. However, sporadic excursions to intermediate states are also seen, and systems become harder and more variable, in parallel with an increase in the fractional contribution of the Comptonization component. These transitions form the HB, and are more frequent in the Cyg-like Z sources than in the Sco-like sources, the latter thought to be intrinsically fainter. 

Again, the most detailed observational evidence for this scenario comes from the analysis of XTE 1701-462. It highest count-rates  correspond to the Cyg-like Z phase, when the system spends most of the time ($\sim 90\%$) between the normal and horizontal ($55\%$ in HB) branches. The latter is particularly interesting since it is associated with an increase in the fast variability, forming an horizontal branch in the RID (right, bottom panel in Fig. \ref{fig:zetas}). Hardness also increases, and thus diagonal tracks are newly seen in the HRD [filled (red) triangles in Fig. \ref{fig:hrd_1701}]. According to the fits presented in LRH09, during the HB the luminosity keeps roughly constant, but thermal emission is converted into Comptonization. A similar behaviour is observed in the bright soft state phases of BHTs, where transitions to the intermediates states (harder and more variable) are seen (e.g. \citealt{Fender2004}, \citealt{Belloni2011}; see also top-left corner in Fig. \ref{fig:gx339} for the case of GX~339-4). Consistently with this, we find that Cyg-like Z sources attain also higher rms levels, as we observe GX~5+1, GX~340+0 and Cyg~X-2 displaying diagonal tracks in their HRD, reaching values in the $10-15\%$ regime, whilst GX~349+2 and GX~17+7 (Sco-like) only show a few scattered points (with rms $<10\%$) above the flat rms hardness relation that characterized the flaring along the bright atoll and weak Z phases.  
\end{enumerate}

\subsubsection{On the absence of bright hard states and hard-to-soft transitions}
A direct consequence of the above scheme is that NS-LMXBs are not seen in the bright hard states ($\lesssim 0.15$ \ledd; see \citealt{Linares2009}), and therefore they do not either perform bright hard-to-soft transitions. Normal atoll phases are observed at luminosities fainter than $\sim 0.3$ \ledd\ (LRH07), their brightest phase corresponding to the soft state, and higher accretion rate sources (BA and Z) sit in thermal dominated states. An exception might be the HB of the Z sources, but their variability properties are only consistent with those observe in the intermediate BH states. 

Luminosities above ($\gtrsim 0.5$ \ledd) are seen in BH systems, especially during the hard-to-soft transitions, but only one system, namely GRS 1915+105, is known to remain persistently at those values (see \citealt{Fender2004a} for a review). Contrastingly to Z sources, it remains most of the time in a hard, high variability state (rms $\sim 20-30 \%$) displaying transitions towards softer states. Other BHs are also seen to transit the area where Z-sources are observed and fast (reverse) transitions from  the bright soft states towards intermediates states are often seen as well. These transitions become less frequent as luminosity decreases through the soft state, which seems to resemble the behaviour of Z sources, where the HB is less frequently formed as the system decays. In Fig. \ref{fig:sketch_BH} we also sketch the behaviour of GRS~1915+105 in the RID on top of that for NS-LMXBs (Fig. \ref{fig:sketch}).

An open question is the luminosity at which the (presumable) hard-to-soft transition took place for XTE~J1701-462. We note that BHTs sometimes display relatively faint hard-soft-transitions, to then become much brighter once the soft state is reached (see red stars in Fig. \ref{fig:gx339}). Given the apparent lack of NS-LMXBs in the bright hard states regions of the digram ($\gtrsim 0.3$ \ledd), this scenario, also observed in NS-LMXBs (e.g. 4U 1705-44; Fig. \ref{h_per}), seem for the moment the most plausible scenario (but unconfirmed) for the rise of the XTE~J1701-462 outburst (or any other transient reaching the bright Z phases).       

\subsubsection{Quasi Periodic Oscillations}
\label{qpos}
In addition to the study of spectral and variability parameters, the detection and evolution of narrow timing features, known as Quasi Periodic Oscillations (QPOs) has proven to be a powerful X-ray state tracker for both BH and NS systems (e.g. \citealt{vanderklis2006}). Three types of QPOs, dubbed A, B and C, are observed in BHTs (see \citealt{Casella2004}; \citealt{Motta2011}). \cite{Casella2005} studied the coincident  evolution of the frequency and the rms of these QPOs (i.e. rms integrated only over the QPO profile), showing similarities between them and those detected in Z sources.  In particular, the A, B, and C QPOs are proposed to be related to the FB, NB and HB oscillations, respectively. Even though a detailed study on this issue is beyond the scope of this paper, it is worth noticing that the proposed scheme results in a good qualitative match if we place both Z and BH QPOs in the RID. However, we note that an exact match between rms values and BH/NS QPO type is not necessarily to be expected since e.g. the rms dynamical range is found to be smaller for NS (up to 50\% for BHs and up to 30\% for NS; see above for possible dilution of the rms due to the NS surface emission).

Type Cs are present during the BH hard states at rms $\gtrsim 10$\%, being particularly strong during the hard-to-soft transitions. We have shown above that the HB covers that region in the RID. Type B and A QPOs are observed in the BH soft states, B observations being slightly more variable. Similarly, FB observations are characterized by variability levels around $~2-3\%$ and NB observations (e.g. low variability inverted triangles in Fig. \ref{fig:hrd_1701}) are also associated with low variability levels, which slightly increase as the system gets close to the HB (LRH09). Atoll QPOs are less well characterized, since they are fainter, but they can be detected during the hard states and the transition to the soft states, showing similar properties to the C oscillations, with frequency increasing with decreasing hardness and rms (e.g. \citealt{vanderklis2006}).\\

Besides to the hard-to-soft transitions of BHTs, the HB can be also compared with the so-called anomalous or ultra-luminous BH state, best observed during the 2005 outburst of the BHT GRO 1655-40 (see \citealt{Belloni2010} and references therein). It occurred once the system reached the soft state, lasted ~20 days and was characterized by significant brightening, spectral hardening and flaring, as well as an increase of the fractional rms up to levels typical of intermediate states (see \citealt{Motta2012} for a full timing analysis). However, we note the evolution of the Type C QPO during this state is odd since frequency do not increase as the system softens (it does the opposite), albeit the spectral analysis is more complex than that during the classical states.   
  
\subsubsection{The accretion-ejection connexion}
\label{radio}

During the last couple of decades, clear connections between the radio and higher energy emission have been established, showing that the X-ray states are tightly connected with ejection phenomena displayed by X-ray binaries (\citealt{Fender2004}; \citealt{Fender2006} for a review). During the BH hard states, a compact jet is seen, and the radio luminosity correlates with the hard (power-law) energy emission (\citealt{Corbel2002}, \citealt{Gallo2003}).  Radio emission starts to fade as the system moves towards the soft state, but close to the state transition, strong radio flares from extended jets moving at relativistic velocities are seen. Once the system reaches the soft state, radio emission is quenched (e.g. \citealt{Russell2011}), and not observed again until the hard states are newly reached at the end of the outburst. A remarkable exception to this are the short harder intervals (i.e. fast transition to the intermediate states) observed during the decay along the bright phases of the soft state. As discussed above, they are associated with an increase in the fast variability, but also to radio the presence of radio flares \citep*{Fender2009}. 

A global study of the radio properties of NS-LMXBs was performed by \cite{Migliari2006}. Even though, radio emission is less powerful in atoll systems, their behaviour roughly follows the above scheme. However, the radio jet is only observed to quench during the bright soft states (above 10\% \ledd; \citealt{Migliari2003}, \citealt{Tudose2009}, \citealt{Miller-Jones2010}), whereas it is still present during the faint soft states. We note that given the known relation between Comptonization and radio emission, this is in agreement with the spectral studies (LRH07) suggesting high Comptonization levels ($\sim 30\%$) during the faint soft states. Our results, consistent with rms decreasing with luminosity during the soft state (see above) also agree with this scenario.\\
Z-sources can be active radio sources, which flaring behaviour resembles that of the persistent black hole GRS 1915+105 (\citealt{Migliari2006}). However, radio emission is not always detected in these systems, and it seems to be associated with transitions from the HB to the NB to then quench in the FB (\citealt{Migliari2006}). Given that the HB coexists in the RID with the hard-to-soft BH transitions, radio flares seem to occur at similar locations for both NSs and BHs. We note however that whereas GRS 1915-105 typically sits in the bright hard state making soft excursions that trigger the radio flaring, Z sources are soft state systems making hard excursions resulting in comparable behaviour (see Fig. \ref{fig:sketch_BH}).

Finally, we note that outflows in the form of accretion disc winds have been observed in BH systems during the phase in which the jet is quenched (i.e. soft states) and vice versa (\citealt{Ponti2012}; see also \citealt{Neilsen2009}). This state-wind connection might also hold for NS-LMXBs \citep*{Ponti2014}, further stressing the similarities between the accretion states of both populations of LMXBs. 

\section{Conclusions}
We have shown that all classes of neutron stars low mass X-ray binaries can be classified following the canonical hard-intermediate-soft state scheme that is now well established for black hole systems. Full transitions between these states result in hysteresis patterns regardless of the transient or persistent nature of the source. Indeed, hysteresis seems to be the natural form that state transitions take below $\sim 30\%$ \ledd. We have shown that hysteresis in neutron stars does not require of dramatic changes in luminosity and even though some differences are found, it follows the same pattern and general properties previously found for black hole transients (see Section \ref{bhvsns}). At higher luminosities hysteresis seems to not be present, but independently of the nature of the compact object, either black hole or neutron star, we have shown that systems located in similar regions of the rms-intensity diagram (i.e. fast variability versus X-ray luminosity) seem to share spectral, timing and multi-wavelength properties, establishing a common framework for accretion states and therefore supporting similar accretion accretion regimes and geometries for both populations of X-ray binaries. Future X-ray spectral and timing, but also multi-wavelength, studies should be able to go deeper into the details of the global scheme presented here.  

\vspace{2cm}

\noindent TMD thanks M. Armas Padilla, M.~ Linares, G.~Ponti and M.~Pretorius for useful comments and discussion. TMD acknowledges funding via an EU Marie Curie Intra-European Fellowship under contract no. 2011-301355. TMD acknowledges hospitality during his visit to ESAC-Madrid and the Instituto de Astrof\'isica de Canarias. TMB acknowledge support from grant INAF PRIN 2012-6. This project was funded in part by European Research Council Advanced Grant 267697 4-pi-sky: Extreme Astrophysics with Revolutionary Radio Telescopes.

\bibliographystyle{mn2e}
\bibliography{/Users/tmd/Dropbox/Libreria.bib} 

\begin{thebibliography}{}

\bibitem[\protect\citeauthoryear{{Armas Padilla}, {Degenaar} \&
  {Wijnands}}{{Armas Padilla} et~al.}{2013}]{ArmasPadilla2013}
{Armas Padilla} M.,  {Degenaar} N.,    {Wijnands} R.,  2013, \mnras, 434, 1586

\bibitem[\protect\citeauthoryear{{Balbus} \& {Henri}}{{Balbus} \&
  {Henri}}{2008}]{Balbus2008}
{Balbus} S.~A.,  {Henri} P.,  2008, \apj, 674, 408

\bibitem[\protect\citeauthoryear{{Barret}}{{Barret}}{2001}]{Barret2001}
{Barret} D.,  2001, Advances in Space Research, 28, 307

\bibitem[\protect\citeauthoryear{{Begelman} \& {Armitage}}{{Begelman} \&
  {Armitage}}{2014}]{Begelman2014}
{Begelman} M.~C.,  {Armitage} P.~J.,  2014, \apjl, 782, L18

\bibitem[\protect\citeauthoryear{{Belloni}, {Homan}, {Casella}, {van der Klis},
  {Nespoli}, {Lewin}, {Miller} \& {M{\'e}ndez}}{{Belloni}
  et~al.}{2005}]{Belloni2005}
{Belloni} T.,  {Homan} J.,  {Casella} P.,  {van der Klis} M.,  {Nespoli} E.,
  {Lewin} W.~H.~G.,  {Miller} J.~M.,    {M{\'e}ndez} M.,  2005, \aap, 440, 207

\bibitem[\protect\citeauthoryear{{Belloni}, {Homan}, {Motta}, {Ratti} \&
  {M{\'e}ndez}}{{Belloni} et~al.}{2007}]{Belloni2007}
{Belloni} T.,  {Homan} J.,  {Motta} S.,  {Ratti} E.,    {M{\'e}ndez} M.,  2007,
  \mnras, 379, 247

\bibitem[\protect\citeauthoryear{{Belloni}, {Parolin}, {Del Santo}, {Homan},
  {Casella}, {Fender}, {Lewin}, {M{\'e}ndez}, {Miller} \& {van der
  Klis}}{{Belloni} et~al.}{2006}]{Belloni2006}
{Belloni} T.,  {Parolin} I.,  {Del Santo} M.,  {Homan} J.,  {Casella} P.,
  {Fender} R.~P.,  {Lewin} W.~H.~G.,  {M{\'e}ndez} M.,  {Miller} J.~M.,    {van
  der Klis} M.,  2006, \mnras, 367, 1113

\bibitem[\protect\citeauthoryear{{Belloni}, {Psaltis} \& {van der
  Klis}}{{Belloni} et~al.}{2002}]{Belloni2002}
{Belloni} T.,  {Psaltis} D.,    {van der Klis} M.,  2002, \apj, 572, 392

\bibitem[\protect\citeauthoryear{{Belloni}}{{Belloni}}{2010}]{Belloni2010}
{Belloni} T.~M.,  2010, Lecture Notes in Physics, Berlin Springer Verlag, 794,
  53

\bibitem[\protect\citeauthoryear{{Belloni}, {Motta} \&
  {Mu{\~n}oz-Darias}}{{Belloni} et~al.}{2011}]{Belloni2011}
{Belloni} T.~M.,  {Motta} S.~E.,    {Mu{\~n}oz-Darias} T.,  2011, Bulletin of
  the Astronomical Society of India, 39, 409

\bibitem[\protect\citeauthoryear{{Capitanio}, {Belloni}, {Del Santo} \&
  {Ubertini}}{{Capitanio} et~al.}{2009}]{Capitanio2009}
{Capitanio} F.,  {Belloni} T.,  {Del Santo} M.,    {Ubertini} P.,  2009,
  \mnras, 398, 1194

\bibitem[\protect\citeauthoryear{{Casella}, {Belloni}, {Homan} \&
  {Stella}}{{Casella} et~al.}{2004}]{Casella2004}
{Casella} P.,  {Belloni} T.,  {Homan} J.,    {Stella} L.,  2004, \aap, 426, 587

\bibitem[\protect\citeauthoryear{{Casella}, {Belloni} \& {Stella}}{{Casella}
  et~al.}{2005}]{Casella2005}
{Casella} P.,  {Belloni} T.,    {Stella} L.,  2005, \apj, 629, 403

\bibitem[\protect\citeauthoryear{{Chenevez}, {Altamirano}, {Galloway}, {in't
  Zand}, {Kuulkers}, {Degenaar}, {Falanga}, {Del Monte}, {Evangelista},
  {Feroci} \& {Costa}}{{Chenevez} et~al.}{2011}]{Chenevez2011}
{Chenevez} J.,  {Altamirano} D.,  {Galloway} D.~K.,  {in't Zand} J.~J.~M.,
  {Kuulkers} E.,  {Degenaar} N.,  {Falanga} M.,  {Del Monte} E.,  {Evangelista}
  Y.,  {Feroci} M.,    {Costa} E.,  2011, \mnras, 410, 179

\bibitem[\protect\citeauthoryear{{Christian} \& {Swank}}{{Christian} \&
  {Swank}}{1997}]{Christian1997}
{Christian} D.~J.,  {Swank} J.~H.,  1997, \apjs, 109, 177

\bibitem[\protect\citeauthoryear{{Church}, {Halai} \&
  {Ba{\l}uci{\'n}ska-Church}}{{Church} et~al.}{2006}]{Church2006}
{Church} M.~J.,  {Halai} G.~S.,    {Ba{\l}uci{\'n}ska-Church} M.,  2006, \aap,
  460, 233

\bibitem[\protect\citeauthoryear{{Corbel}, {Fender}, {Tzioumis}, {Tomsick},
  {Orosz}, {Miller}, {Wijnands} \& {Kaaret}}{{Corbel}
  et~al.}{2002}]{Corbel2002}
{Corbel} S.,  {Fender} R.~P.,  {Tzioumis} A.~K.,  {Tomsick} J.~A.,  {Orosz}
  J.~A.,  {Miller} J.~M.,  {Wijnands} R.,    {Kaaret} P.,  2002, Science, 298,
  196

\bibitem[\protect\citeauthoryear{{di Salvo}, {D'A{\'{\i}}}, {Iaria}, {Burderi},
  {Dov{\v c}iak}, {Karas}, {Matt}, {Papitto}, {Piraino}, {Riggio}, {Robba} \&
  {Santangelo}}{{di Salvo} et~al.}{2009}]{diSalvo2009}
{di Salvo} T.,  {D'A{\'{\i}}} A.,  {Iaria} R.,  {Burderi} L.,  {Dov{\v c}iak}
  M.,  {Karas} V.,  {Matt} G.,  {Papitto} A.,  {Piraino} S.,  {Riggio} A.,
  {Robba} N.~R.,    {Santangelo} A.,  2009, \mnras, 398, 2022

\bibitem[\protect\citeauthoryear{{Di Salvo}, {Iaria}, {Burderi} \& {Robba}}{{Di
  Salvo} et~al.}{2000}]{diSalvo2000}
{Di Salvo} T.,  {Iaria} R.,  {Burderi} L.,    {Robba} N.~R.,  2000, \apj, 542,
  1034

\bibitem[\protect\citeauthoryear{{D{\'{\i}}az Trigo}, {Boirin}, {Costantini},
  {M{\'e}ndez} \& {Parmar}}{{D{\'{\i}}az Trigo} et~al.}{2011}]{DiazTrigo2011}
{D{\'{\i}}az Trigo} M.,  {Boirin} L.,  {Costantini} E.,  {M{\'e}ndez} M.,
  {Parmar} A.,  2011, \aap, 528, A150

\bibitem[\protect\citeauthoryear{{Done} \& {Gierli{\'n}ski}}{{Done} \&
  {Gierli{\'n}ski}}{2003}]{Done2003}
{Done} C.,  {Gierli{\'n}ski} M.,  2003, \mnras, 342, 1041

\bibitem[\protect\citeauthoryear{{Dunn}, {Fender}, {K{\"o}rding}, {Belloni} \&
  {Cabanac}}{{Dunn} et~al.}{2010}]{Dunn2010}
{Dunn} R.~J.~H.,  {Fender} R.~P.,  {K{\"o}rding} E.~G.,  {Belloni} T.,
  {Cabanac} C.,  2010, \mnras, 403, 61

\bibitem[\protect\citeauthoryear{{Fender}}{{Fender}}{2006}]{Fender2006}
{Fender} R.,  2006, in Compact stellar X-ray sources, pp 381--419

\bibitem[\protect\citeauthoryear{{Fender} \& {Belloni}}{{Fender} \&
  {Belloni}}{2004}]{Fender2004a}
{Fender} R.,  {Belloni} T.,  2004, \araa, 42, 317

\bibitem[\protect\citeauthoryear{{Fender}, {Belloni} \& {Gallo}}{{Fender}
  et~al.}{2004}]{Fender2004}
{Fender} R.~P.,  {Belloni} T.~M.,    {Gallo} E.,  2004, \mnras, 355, 1105

\bibitem[\protect\citeauthoryear{{Fender}, {Dahlem}, {Homan}, {Corbel}, {Sault}
  \& {Belloni}}{{Fender} et~al.}{2007}]{Fender2007}
{Fender} R.~P.,  {Dahlem} M.,  {Homan} J.,  {Corbel} S.,  {Sault} R.,
  {Belloni} T.~M.,  2007, \mnras, 380, L25

\bibitem[\protect\citeauthoryear{{Fender}, {Homan} \& {Belloni}}{{Fender}
  et~al.}{2009}]{Fender2009}
{Fender} R.~P.,  {Homan} J.,    {Belloni} T.~M.,  2009, \mnras, 396, 1370

\bibitem[\protect\citeauthoryear{{Gallo}, {Fender} \& {Pooley}}{{Gallo}
  et~al.}{2003}]{Gallo2003}
{Gallo} E.,  {Fender} R.~P.,    {Pooley} G.~G.,  2003, \mnras, 344, 60

\bibitem[\protect\citeauthoryear{{Galloway}, {Muno}, {Hartman}, {Psaltis} \&
  {Chakrabarty}}{{Galloway} et~al.}{2008}]{Galloway2008}
{Galloway} D.~K.,  {Muno} M.~P.,  {Hartman} J.~M.,  {Psaltis} D.,
  {Chakrabarty} D.,  2008, accepted by ApJS (Astro-ph 0608259)

\bibitem[\protect\citeauthoryear{{Gierli{\'n}ski}, {Zdziarski}, {Poutanen},
  {Coppi}, {Ebisawa} \& {Johnson}}{{Gierli{\'n}ski}
  et~al.}{1999}]{Gierlinski1999}
{Gierli{\'n}ski} M.,  {Zdziarski} A.~A.,  {Poutanen} J.,  {Coppi} P.~S.,
  {Ebisawa} K.,    {Johnson} W.~N.,  1999, \mnras, 309, 496

\bibitem[\protect\citeauthoryear{{Gilfanov}}{{Gilfanov}}{2010}]{Gilfanov2010}
{Gilfanov} M.,  2010, in {T.~Belloni} ed., Lecture Notes in Physics, Berlin
  Springer Verlag Vol.~794 of Lecture Notes in Physics, Berlin Springer Verlag,
  {X-Ray Emission from Black-Hole Binaries}.
pp 17--+

\bibitem[\protect\citeauthoryear{{Gladstone}, {Done} \&
  {Gierli{\'n}ski}}{{Gladstone} et~al.}{2007}]{Gladstone2007}
{Gladstone} J.,  {Done} C.,    {Gierli{\'n}ski} M.,  2007, \mnras, 378, 13

\bibitem[\protect\citeauthoryear{{Hasinger} \& {van der Klis}}{{Hasinger} \&
  {van der Klis}}{1989}]{Hasinger1989}
{Hasinger} G.,  {van der Klis} M.,  1989, \aap, 225, 79

\bibitem[\protect\citeauthoryear{{Heil}, {Vaughan} \& {Uttley}}{{Heil}
  et~al.}{2012}]{Heil2012}
{Heil} L.~M.,  {Vaughan} S.,    {Uttley} P.,  2012, \mnras, 422, 2620

\bibitem[\protect\citeauthoryear{{Heinke}, {Edmonds}, {Grindlay}, {Lloyd},
  {Cohn} \& {Lugger}}{{Heinke} et~al.}{2003}]{Heinke2003}
{Heinke} C.~O.,  {Edmonds} P.~D.,  {Grindlay} J.~E.,  {Lloyd} D.~A.,  {Cohn}
  H.~N.,    {Lugger} P.~M.,  2003, \apj, 590, 809

\bibitem[\protect\citeauthoryear{{Homan}, {van der Klis}, {Fridriksson},
  {Remillard}, {Wijnands}, {M{\'e}ndez}, {Lin}, {Altamirano}, {Casella},
  {Belloni} \& {Lewin}}{{Homan} et~al.}{2010}]{Homan2010}
{Homan} J.,  {van der Klis} M.,  {Fridriksson} J.~K.,  {Remillard} R.~A.,
  {Wijnands} R.,  {M{\'e}ndez} M.,  {Lin} D.,  {Altamirano} D.,  {Casella} P.,
  {Belloni} T.~M.,    {Lewin} W.~H.~G.,  2010, \apj, 719, 201

\bibitem[\protect\citeauthoryear{{Homan}, {van der Klis}, {Wijnands},
  {Belloni}, {Fender}, {Klein-Wolt}, {Casella}, {M{\'e}ndez}, {Gallo}, {Lewin}
  \& {Gehrels}}{{Homan} et~al.}{2007}]{Homan2007}
{Homan} J.,  {van der Klis} M.,  {Wijnands} R.,  {Belloni} T.,  {Fender} R.,
  {Klein-Wolt} M.,  {Casella} P.,  {M{\'e}ndez} M.,  {Gallo} E.,  {Lewin}
  W.~H.~G.,    {Gehrels} N.,  2007, \apj, 656, 420

\bibitem[\protect\citeauthoryear{{Homan}, {Wijnands}, {van der Klis},
  {Belloni}, {van Paradijs}, {Klein-Wolt}, {Fender} \& {M\'endez}}{{Homan}
  et~al.}{2001}]{Homan2001}
{Homan} J.,  {Wijnands} R.,  {van der Klis} M.,  {Belloni} T.,  {van Paradijs}
  J.,  {Klein-Wolt} M.,  {Fender} R.,    {M\'endez} M.,  2001, \apjs, 132, 377

\bibitem[\protect\citeauthoryear{{Kuulkers}, {van der Klis}, {Oosterbroek},
  {Asai}, {Dotani}, {van Paradijs} \& {Lewin}}{{Kuulkers}
  et~al.}{1994}]{Kuulkers1994}
{Kuulkers} E.,  {van der Klis} M.,  {Oosterbroek} T.,  {Asai} K.,  {Dotani} T.,
   {van Paradijs} J.,    {Lewin} W.~H.~G.,  1994, \aap, 289, 795

\bibitem[\protect\citeauthoryear{{Li}, {Song}, {Qu}, {Lei}, {Nie} \&
  {Zhang}}{{Li} et~al.}{2012}]{Li2012}
{Li} Z.~B.,  {Song} L.~M.,  {Qu} J.~L.,  {Lei} Y.~J.,  {Nie} J.~Y.,    {Zhang}
  C.~M.,  2012, \apss, 341, 383

\bibitem[\protect\citeauthoryear{{Lin}, {Altamirano}, {Homan}, {Remillard},
  {Wijnands} \& {Belloni}}{{Lin} et~al.}{2009b}]{Lin2009}
{Lin} D.,  {Altamirano} D.,  {Homan} J.,  {Remillard} R.~A.,  {Wijnands} R.,
  {Belloni} T.,  2009b, \apj, 699, 60

\bibitem[\protect\citeauthoryear{{Lin}, {Remillard} \& {Homan}}{{Lin}
  et~al.}{2007}]{Lin2007}
{Lin} D.,  {Remillard} R.~A.,    {Homan} J.,  2007, \apj, 667, 1073

\bibitem[\protect\citeauthoryear{{Lin}, {Remillard} \& {Homan}}{{Lin}
  et~al.}{2009}]{Lin2009a}
{Lin} D.,  {Remillard} R.~A.,    {Homan} J.,  2009, \apj, 696, 1257

\bibitem[\protect\citeauthoryear{{Linares}}{{Linares}}{2009}]{Linares2009}
{Linares} M.,  2009, PhD thesis, Sterrenkundig Instituut ''Anton Pannekoek -
  University of Amsterdam

\bibitem[\protect\citeauthoryear{{Maccarone}}{{Maccarone}}{2003}]{Maccarone200%
3}
{Maccarone} T.~J.,  2003, \aap, 409, 697

\bibitem[\protect\citeauthoryear{{Maccarone} \& {Coppi}}{{Maccarone} \&
  {Coppi}}{2003}]{Maccarone2003a}
{Maccarone} T.~J.,  {Coppi} P.~S.,  2003, \mnras, 338, 189

\bibitem[\protect\citeauthoryear{{Migliari} \& {Fender}}{{Migliari} \&
  {Fender}}{2006}]{Migliari2006}
{Migliari} S.,  {Fender} R.~P.,  2006, \mnras, 366, 79

\bibitem[\protect\citeauthoryear{{Migliari}, {Fender}, {Rupen}, {Jonker},
  {Klein-Wolt}, {Hjellming} \& {van der Klis}}{{Migliari}
  et~al.}{2003}]{Migliari2003}
{Migliari} S.,  {Fender} R.~P.,  {Rupen} M.,  {Jonker} P.~G.,  {Klein-Wolt} M.,
   {Hjellming} R.~M.,    {van der Klis} M.,  2003, \mnras, 342, L67

\bibitem[\protect\citeauthoryear{{Miller-Jones}, {Sivakoff}, {Altamirano},
  {Tudose}, {Migliari}, {Dhawan}, {Fender}, {Garrett}, {Heinz}, {K{\"o}rding},
  {Krimm}, {Linares}, {Maitra}, {Markoff}, {Paragi} \&
  {Remillard}}{{Miller-Jones} et~al.}{2010}]{Miller-Jones2010}
{Miller-Jones} J.~C.~A.,  {Sivakoff} G.~R.,  {Altamirano} D.,  {Tudose} V.,
  {Migliari} S.,  {Dhawan} V.,  {Fender} R.~P.,  {Garrett} M.~A.,  {Heinz} S.,
  {K{\"o}rding} E.~G.,  {Krimm} H.~A.,  {Linares} M.,  {Maitra} D.,  {Markoff}
  S.,  {Paragi} Z.,    {Remillard} R.~A. e.~a.,  2010, \apjl, 716, L109

\bibitem[\protect\citeauthoryear{{Mitsuda}, {Inoue}, {Nakamura} \&
  {Tanaka}}{{Mitsuda} et~al.}{1989}]{Mitsuda1989}
{Mitsuda} K.,  {Inoue} H.,  {Nakamura} N.,    {Tanaka} Y.,  1989, \pasj, 41, 97

\bibitem[\protect\citeauthoryear{{Miyamoto}, {Iga}, {Kitamoto} \&
  {Kamado}}{{Miyamoto} et~al.}{1993}]{Miyamoto1993}
{Miyamoto} S.,  {Iga} S.,  {Kitamoto} S.,    {Kamado} Y.,  1993, \apjl, 403,
  L39

\bibitem[\protect\citeauthoryear{{Miyamoto}, {Kitamoto}, {Hayashida} \&
  {Egoshi}}{{Miyamoto} et~al.}{1995}]{Miyamoto1995}
{Miyamoto} S.,  {Kitamoto} S.,  {Hayashida} K.,    {Egoshi} W.,  1995, \apjl,
  442, L13

\bibitem[\protect\citeauthoryear{{Miyamoto}, {Kitamoto}, {Iga}, {Negoro} \&
  {Terada}}{{Miyamoto} et~al.}{1992}]{Miyamoto1992}
{Miyamoto} S.,  {Kitamoto} S.,  {Iga} S.,  {Negoro} H.,    {Terada} K.,  1992,
  \apjl, 391, L21

\bibitem[\protect\citeauthoryear{{Motta}, {Belloni} \& {Homan}}{{Motta}
  et~al.}{2009}]{Motta2009}
{Motta} S.,  {Belloni} T.,    {Homan} J.,  2009, \mnras, 400, 1603

\bibitem[\protect\citeauthoryear{{Motta}, {Homan}, {Mu{\~n}oz Darias},
  {Casella}, {Belloni}, {Hiemstra} \& {M{\'e}ndez}}{{Motta}
  et~al.}{2012}]{Motta2012}
{Motta} S.,  {Homan} J.,  {Mu{\~n}oz Darias} T.,  {Casella} P.,  {Belloni}
  T.~M.,  {Hiemstra} B.,    {M{\'e}ndez} M.,  2012, \mnras, 427, 595

\bibitem[\protect\citeauthoryear{{Motta}, {Mu{\~n}oz-Darias} \&
  {Belloni}}{{Motta} et~al.}{2010}]{Motta2010}
{Motta} S.,  {Mu{\~n}oz-Darias} T.,    {Belloni} T.,  2010, \mnras, 408, 1796

\bibitem[\protect\citeauthoryear{{Motta}, {Mu{\~n}oz-Darias}, {Casella},
  {Belloni} \& {Homan}}{{Motta} et~al.}{2011}]{Motta2011}
{Motta} S.,  {Mu{\~n}oz-Darias} T.,  {Casella} P.,  {Belloni} T.,    {Homan}
  J.,  2011, \mnras, 418, 2292

\bibitem[\protect\citeauthoryear{{Mu{\~n}oz-Darias}, {Coriat}, {Plant},
  {Ponti}, {Fender} \& {Dunn}}{{Mu{\~n}oz-Darias}
  et~al.}{2013}]{Munoz-Darias2013b}
{Mu{\~n}oz-Darias} T.,  {Coriat} M.,  {Plant} D.~S.,  {Ponti} G.,  {Fender}
  R.~P.,    {Dunn} R.~J.~H.,  2013, \mnras

\bibitem[\protect\citeauthoryear{{Mu{\~n}oz-Darias}, {Motta} \&
  {Belloni}}{{Mu{\~n}oz-Darias} et~al.}{2011}]{Munoz-Darias2011}
{Mu{\~n}oz-Darias} T.,  {Motta} S.,    {Belloni} T.~M.,  2011, \mnras, 410, 679

\bibitem[\protect\citeauthoryear{{Mu{\~n}oz-Darias}, {Motta}, {Pawar},
  {Belloni}, {Campana} \& {Bhattacharya}}{{Mu{\~n}oz-Darias}
  et~al.}{2010}]{Munoz-Darias2010}
{Mu{\~n}oz-Darias} T.,  {Motta} S.,  {Pawar} D.,  {Belloni} T.~M.,  {Campana}
  S.,    {Bhattacharya} D.,  2010, \mnras, 404, L94

\bibitem[\protect\citeauthoryear{{Mu{\~n}oz-Darias}, {Motta}, {Stiele} \&
  {Belloni}}{{Mu{\~n}oz-Darias} et~al.}{2011}]{Munoz-Darias2011b}
{Mu{\~n}oz-Darias} T.,  {Motta} S.,  {Stiele} H.,    {Belloni} T.~M.,  2011b,
  \mnras, 415, 292

\bibitem[\protect\citeauthoryear{{Neilsen} \& {Lee}}{{Neilsen} \&
  {Lee}}{2009}]{Neilsen2009}
{Neilsen} J.,  {Lee} J.~C.,  2009, \nat, 458, 481

\bibitem[\protect\citeauthoryear{{Nixon} \& {Salvesen}}{{Nixon} \&
  {Salvesen}}{2014}]{Nixon2014}
{Nixon} C.,  {Salvesen} G.,  2014, \mnras, 437, 3994

\bibitem[\protect\citeauthoryear{{Petrucci}, {Ferreira}, {Henri} \&
  {Pelletier}}{{Petrucci} et~al.}{2008}]{Petrucci2008}
{Petrucci} P.-O.,  {Ferreira} J.,  {Henri} G.,    {Pelletier} G.,  2008,
  \mnras, 385, L88

\bibitem[\protect\citeauthoryear{{Ponti}, {Fender}, {Begelman}, {Dunn},
  {Neilsen} \& {Coriat}}{{Ponti} et~al.}{2012}]{Ponti2012}
{Ponti} G.,  {Fender} R.~P.,  {Begelman} M.~C.,  {Dunn} R.~J.~H.,  {Neilsen}
  J.,    {Coriat} M.,  2012, \mnras, 422, L11

\bibitem[\protect\citeauthoryear{{Ponti}, {Mu{\~n}oz-Darias} \&
  {Fender}}{{Ponti} et~al.}{2014}]{Ponti2014}
{Ponti} G.,  {Mu{\~n}oz-Darias} T.,    {Fender} R.~P.,  2014, Revised version
  submitted to \mnras

\bibitem[\protect\citeauthoryear{{Remillard}, {Lin}, {Cooper} \&
  {Narayan}}{{Remillard} et~al.}{2006}]{Remillard2006}
{Remillard} R.~A.,  {Lin} D.,  {Cooper} R.~L.,    {Narayan} R.,  2006, \apj,
  646, 407

\bibitem[\protect\citeauthoryear{{Russell}, {Miller-Jones}, {Maccarone},
  {Yang}, {Fender} \& {Lewis}}{{Russell} et~al.}{2011}]{Russell2011}
{Russell} D.~M.,  {Miller-Jones} J.~C.~A.,  {Maccarone} T.~J.,  {Yang} Y.~J.,
  {Fender} R.~P.,    {Lewis} F.,  2011, \apjl, 739, L19

\bibitem[\protect\citeauthoryear{{Sanna}, {M{\'e}ndez}, {Altamirano}, {Homan},
  {Casella}, {Belloni}, {Lin}, {van der Klis} \& {Wijnands}}{{Sanna}
  et~al.}{2010}]{Sanna2010}
{Sanna} A.,  {M{\'e}ndez} M.,  {Altamirano} D.,  {Homan} J.,  {Casella} P.,
  {Belloni} T.,  {Lin} D.,  {van der Klis} M.,    {Wijnands} R.,  2010, \mnras,
  408, 622

\bibitem[\protect\citeauthoryear{{Savolainen}, {Hannikainen}, {Vilhu},
  {Paizis}, {Nevalainen} \& {Hakala}}{{Savolainen}
  et~al.}{2009}]{Savolainen2009}
{Savolainen} P.,  {Hannikainen} D.~C.,  {Vilhu} O.,  {Paizis} A.,  {Nevalainen}
  J.,    {Hakala} P.,  2009, \mnras, 393, 569

\bibitem[\protect\citeauthoryear{{Shakura} \& {Sunyaev}}{{Shakura} \&
  {Sunyaev}}{1973}]{Shakura1973}
{Shakura} N.~I.,  {Sunyaev} R.~A.,  1973, \aap, 24, 337

\bibitem[\protect\citeauthoryear{{Shaw}, {Charles}, {Bird}, {Cornelisse},
  {Casares}, {Lewis}, {Mu{\~n}oz-Darias}, {Russell} \& {Zurita}}{{Shaw}
  et~al.}{2013}]{shaw2013}
{Shaw} A.~W.,  {Charles} P.~A.,  {Bird} A.~J.,  {Cornelisse} R.,  {Casares} J.,
   {Lewis} F.,  {Mu{\~n}oz-Darias} T.,  {Russell} D.~M.,    {Zurita} C.,  2013,
  \mnras, 433, 740

\bibitem[\protect\citeauthoryear{{Soleri}, {Mu{\~n}oz-Darias}, {Motta},
  {Belloni}, {Casella}, {M{\'e}ndez}, {Altamirano}, {Linares}, {Wijnands},
  {Fender} \& {van der Klis}}{{Soleri} et~al.}{2013}]{Soleri2013}
{Soleri} P.,  {Mu{\~n}oz-Darias} T.,  {Motta} S.,  {Belloni} T.,  {Casella} P.,
   {M{\'e}ndez} M.,  {Altamirano} D.,  {Linares} M.,  {Wijnands} R.,  {Fender}
  R.,    {van der Klis} M.,  2013, \mnras, 429, 1244

\bibitem[\protect\citeauthoryear{{Stiele}, {Motta}, {Mu{\~n}oz-Darias} \&
  {Belloni}}{{Stiele} et~al.}{2011}]{Stiele2011}
{Stiele} H.,  {Motta} S.,  {Mu{\~n}oz-Darias} T.,    {Belloni} T.~M.,  2011,
  \mnras, 418, 1746

\bibitem[\protect\citeauthoryear{{Stiele}, {Mu{\~n}oz-Darias}, {Motta} \&
  {Belloni}}{{Stiele} et~al.}{2012}]{Stiele2012}
{Stiele} H.,  {Mu{\~n}oz-Darias} T.,  {Motta} S.,    {Belloni} T.~M.,  2012,
  \mnras, 422, 679

\bibitem[\protect\citeauthoryear{{Syunyaev} \& {Shakura}}{{Syunyaev} \&
  {Shakura}}{1986}]{Syunyaev1986}
{Syunyaev} R.~A.,  {Shakura} N.~I.,  1986, Soviet Astronomy Letters, 12, 117

\bibitem[\protect\citeauthoryear{{Tudose}, {Fender}, {Linares}, {Maitra} \&
  {van der Klis}}{{Tudose} et~al.}{2009}]{Tudose2009}
{Tudose} V.,  {Fender} R.~P.,  {Linares} M.,  {Maitra} D.,    {van der Klis}
  M.,  2009, \mnras, 400, 2111

\bibitem[\protect\citeauthoryear{{van der Klis}}{{van der
  Klis}}{2006}]{vanderklis2006}
{van der Klis} M.,  2006, pp 39--112

\bibitem[\protect\citeauthoryear{{White}, {Stella} \& {Parmar}}{{White}
  et~al.}{1988}]{White1988}
{White} N.~E.,  {Stella} L.,    {Parmar} A.~N.,  1988, \apj, 324, 363

\bibitem[\protect\citeauthoryear{{Wijnands} \& {van der Klis}}{{Wijnands} \&
  {van der Klis}}{1999}]{Wijnands1999a}
{Wijnands} R.,  {van der Klis} M.,  1999, \apj, 514, 939

\bibitem[\protect\citeauthoryear{{Wijnands}, {van der Klis}, {Kuulkers}, {Asai}
  \& {Hasinger}}{{Wijnands} et~al.}{1997}]{Wijnands1997}
{Wijnands} R.~A.~D.,  {van der Klis} M.,  {Kuulkers} E.,  {Asai} K.,
  {Hasinger} G.,  1997, \aap, 323, 399

\bibitem[\protect\citeauthoryear{{Wilms}, {Nowak}, {Pottschmidt}, {Pooley} \&
  {Fritz}}{{Wilms} et~al.}{2006}]{Wilms2006}
{Wilms} J.,  {Nowak} M.~A.,  {Pottschmidt} K.,  {Pooley} G.~G.,    {Fritz} S.,
  2006, \aap, 447, 245

\end{thebibliography}
\nocite{Belloni2010}
\nocite{Munoz-Darias2011b}
\nocite{Lin2009}
\nocite{Lin2007}
\nocite{Miyamoto1993}
\label{lastpage}

\appendix
\section{Full list of sources}
\vspace{-2 cm}
\begin{table*}
\centering
\caption{List of the 50 sources analysed for this work. Comments roughly describe the behaviour observed in the HIDs, RIDs and HRDs obtained from the RXTE/PCU observations analysed in this work.}

\begin{tabular}{ l c l }
\hline
\textbf{NS-LMXB} & \textbf{Number of Observations analysed} & \textbf{Comments}\\
\hline
IGR~J00291+5934 & 53 & Transient. Hard State \\
4U~0513-40 & 320 & Persistent. Hysteresis-like behaviour? (low count rates) \\
LMC~X-2 (4U~0520-72) & 131 & Persistent. Bright soft state \\
MAXI~J0556-332 & 262 & Transient. Probably decaying along the bright soft state (including harder intervals) \\
4U~0614+09 & 474 & Persistent. Hysteresis loops \\
EXO~0748-676 & 707 & Long-term transient. Hysteresis loops \\
4U~0918-549 & 243 & Persistent. Hard state \\
XTE~J0929-314 & 34 & Transient. Hard State  \\ 
XB~1254-690 & 90 & Persistent. Bright soft state \\
XB~1323-619 & 27 & Persistent. Hard state \\
4U~1608-52 & 980 & Transient. Hysteresis loops \\
4U 1624-490 & 21 & Persistent. Soft state\\
4U~1636-536 & 1414 & Persistent. Hysteresis loops \\
GX~340+0 (4U~1642-45) & 90 & Persistent. Bright soft state (including harder intervals) \\
MXB~1659-298 & 62 & Transient. Soft state \\
XTE~J1701-462 & 835 & Transient. All the states\\
GX~349+2 (4U~1702-36) & 134 & Persistent. Bright soft state (including harder intervals) \\
4U~1702-429 & 116 & Persistent. Hysteresis-like behaviour\\
4U~1705-44 & 506 & Persistent. Hysteresis loops \\
XTE~J1709-267 & 36 & Transient. Hysteresis-like behaviour \\
XTE~J1710-281 & 27 & Transient. Hard state\\
4U~1724-30 & 117 & Persistent. Hysteresis-like behaviour \\
GX~9+9 (4U~1728-16 )& 71 & Persistent. Bright soft state \\
4U~1728-34 & 356 & Persistent. Hysteresis loops \\
KS~1731-260 & 68 & Long-term transient. Hysteresis-like behaviour \\
4U~1735-44 & 210 & Persistent. Soft state - bright soft state \\ 
SLX~1735-269 & 64 &  Persistent. Hysteresis-like behaviour\\
GX~3+1 (XB~1744-265) & 98 & Persistent. Bright soft state \\
1A~1744-361 & 47 & Transient. Mostly soft. A few hard observations at low rates \\
EXO~1745-248  & 33 & Transient. Hysteresis loops \\
4U~1746-371 & 49 & Persistent. Mostly soft. A few hard observations at low rates \\
IGR~17473-2721 & 179 & Transient. Hysteresis loops \\
IGR~J17480-2446 & 46 & Transient. Mostly soft.\\
XTE~J1751-305 & 75 & Transient. Hard State \\
GX~9+1 (4U~1758-20) & 91 & Persistent. Bright soft state \\
GX~5-1 (4U~1758-25) & 159 & Persistent. Bright soft state (including harder intervals) \\
XTE~J1759-220 & 36 & Long-term transient. Hysteresis-like behaviour \\
XTE~J1807-294 & 83 & Transient. Hard State \\
SAX~J1808.4-3658 & 246 & Transient. Hard State \\
GX~13+1 (4U~1811-17)  & 82 & Persistent. Bright soft state \\
GX~17+2 (4U~1813-14) & 177 & Persistent. Bright soft state (including harder intervals) \\
XTE~J1814-338 & 58 & Transient. Hard State \\
4U~1820-303 & 208 & Persistent. Hysteresis (1 loop) \\
X~1822-371 & 89 & Persistent (Accretion disc corona). Bright soft state (harder intervals?) \\
GS~1826-24 & 135 & Persistent (or long-term transient). Hard State \\
Ser~X-1 (4U 1837+04) & 94 & Persistent. Bright soft state \\
HETE~J1900.1-2455 & 344 & Long-Term transient. Mostly hard. Softer intervals seem to follow hysteresis patterns.\\
Aql~X-1 (4U~1908+005) & 465 & Transient. Hysteresis loops \\
4U~1916-053 & 43 & Persistent. Mostly hard with some softer observations\\
Cyg~X-2 (4U~2142+38) & 531 & Persistent. Bright soft state (including harder intervals) \\
\hline
\end{tabular}
\label{tab:full-list}
\end{table*}  

\end{document}